# Statistical characterization of pulsar glitches and their potential impact on searches for continuous gravitational waves

G. Ashton,[1,2,*] R. Prix,[1] and D. I. Jones[2]

[1]*Max Planck Institut für Gravitationsphysik (Albert Einstein Institut) and Leibniz Universität Hannover, 30161 Hannover, Germany*
[2]*Mathematical Sciences, University of Southampton, Southampton SO17 1BJ, United Kingdom*

Continuous gravitational waves from neutron stars could provide an invaluable resource to learn about their interior physics. A common search method involves matched filtering a modeled template against the noisy gravitational-wave data to find signals. This method suffers a *mismatch* (i.e., relative loss of the signal-to-noise ratio) if the signal deviates from the template. One possible instance in which this may occur is if the neutron star undergoes a *glitch*, a sudden rapid increase in the rotation frequency seen in the timing of many radio pulsars. In this work, we use a statistical characterization of the glitch rate and size in radio pulsars to estimate how often neutron star glitches would occur within the parameter space of continuous gravitational-wave searches and how much mismatch putative signals would suffer in the search due to these glitches. We find that for many previous and potential future searches continuous-wave signals have an elevated probability of undergoing one or more glitches and that these glitches will often lead to a substantial fraction of the signal-to-noise ratio being lost. This could lead to a failure to identify candidate gravitational-wave signals in the initial stages of a search and also to the false dismissal of candidates in subsequent follow-up stages.

## I. INTRODUCTION

Electromagnetic (EM) observations of pulsar glitches have long been one of the most fruitful sources of insight into neutron star physics. They are characterized by a sudden increase[1] in the rotation frequency, often accompanied by a jump in the frequency derivative and an exponential recovery of some fraction of the initial frequency jump. The events happen rapidly and are sufficiently disruptive that pulsar timing models often lose phase coherence over the event.

Two leading models exist to explain glitches. In the *superfluid pinning* model, some portion of the interior superfluid is pinned and does not participate in the smooth torque-driven spin-down of the rest of the crust (where "crust" refers to the actual crust, plus whatever other parts of the star that are strongly coupled to it). After some period, the crust will therefore have developed a frequency lag compared to the pinned superfluid. A glitch occurs when the two components recouple, transferring angular momentum from the pinned superfluid to the crust and producing a spin-up of the crust [2,3]. Alternatively, glitches could be caused by *crust cracking* as the crust readjusts to a minimum energy configuration brought about by the gradual decay of the spin-down rate [4]. It is also possible that glitches result from a combination of these two models. In either case, it seems reasonable to assume that both the crust and the core will be involved.

Rotating isolated neutron stars can produce continuous gravitational waves (CWs) from nonaxisymmetric distortions. These distortions can be stationary as viewed from the frame rotating with the star, so-called "mountains" (supported by either elastic stresses in the crust or by magnetic fields), or persistent oscillation modes of the neutron star. In the case of mountains, the star emits a CW at a frequency $f_s$ that is twice the rotation frequency $\nu$, i.e., $f_s = 2\nu$. In the case of oscillation modes, the frequency relation will be different and generally depends on the equation of state and the type of oscillation (e.g., $f_s \approx 4\nu/3$ in the case of r modes; see the review by Prix [5] for an overview of different CW emission mechanisms). In the following, we assume the mountain model for simplicity.

It is possible (see, for example, the work by van Eysden and Melatos [6] and Keer and Jones [7], or Singh [8]) that a glitch could trigger a quadrupolar quasinormal mode resulting in a transient burst of gravitational waves; search methodologies for such a signal have been considered by Clark *et al.* [9] for signals lasting $\mathcal{O}(\mathrm{ms})$ and by Prix *et al.* [10] for signals lasting $\mathcal{O}(\mathrm{hours-weeks})$. However, in this work, we are not concerned with the gravitational radiation triggered by a glitch but rather the impact glitches may have on searches for CW signals. Specifically, assuming an isolated rotating neutron star is producing a CW signal, if a glitch occurs and causes a sudden increase in the CW frequency, what will the impact be on our ability to detect the signal?

---

[*]gregory.ashton@ligo.org

[1]A few cases of "antiglitches," in which the frequency decreases, have also been observed (e.g., the work of Archibald *et al.* [1]); however, these constitute a minority of all observed glitches, and so we will not consider them in this work.

Estimates for the intrinsic gravitational-wave strain amplitude $h_0$ for canonical models of CW emissions (see, for example, the work of Abbott *et al.* [11]) suggest they are extremely weak compared to the noise level of advanced detectors [12]. To detect a signal, significant effort has been put into data analysis methods, which may be capable of identifying the putative signals. Many of these methods rely on *matched filtering* in which a template is correlated with the data in the hope of detecting the presence of the unknown signal similar to the template. The power of these methods lies in the fact that the signal-to-noise ratio (SNR) grows as the square root of the observation time (e.g., see Ref. [5] for an overview). Because of the longevity of CW signals, this allows the weak signal to be discerned from the noise by using a sufficiently long stretch of data.

These methods are powerful but harbor a vulnerability in any instance in which the signal lies outside of the regular CW template manifold (as defined in Sec. IV A). For CW signals from nonaxisymmetric distortions of neutron stars, we can expect that discrepancies from such a template may manifest in one of two ways. First, we know from radio pulsar timing that the spin-down of a pulsar differs from a smooth spin-down due to *timing noise*. This is a continuous low-frequency structure in the residual between the best-fit Taylor-series timing model and the observed pulsations (for a review, see the work of Hobbs *et al.* [13]). The effect of timing noise on CW searches was studied by Jones [14] and Ashton *et al.* [15], where it was found that its presence limits the coherent observation span over which signals may be detected if searched for using a smooth signal. However, one can attempt to mitigate this effect when performing a targeted search for a known radio pulsar by including the timing noise seen in the EM channel in the template; a method to do this is described by Pitkin and Woan [16]. In this work, we will address the second potential discrepancy between the signal and template: glitches.

There are two distinct questions to answer in the case of glitches:
  (1) How probable is it that a glitch will occur during our CW observation?
  (2) If a glitch does occur, what effect will it have on our ability to discover the CW signal?

To answer these questions, we use known radio pulsar glitch statistics to estimate the size and rate of radio pulsar glitches for the parameter spaces considered in typical CW searches. In other words, we assume that the glitches we observe in the radio pulsars population are representative of those that we may expect to see in the population of CW signals. This is purely pragmatic in that we do not know of any other populations on which to base our assumptions. We then quantify the effect such glitches will have on current CW detection methods by calculating the mismatch, i.e., the relative loss of the squared signal-to-noise ratio. We do this by modeling a glitch as a piecewise Taylor-series expansion with a discontinuity at the glitch; we do not model the exponential recovery observed in some glitches, but we will discuss the significance this may have in Sec. V C. Ultimately, the goal of this work is to estimate the risk faced by current and ongoing CW searches to glitches in their target population.

In Sec. II, we will briefly describe current CW searches and how glitches may effect them. Then, in Sec. III, we investigate the statistical properties of the observed radio pulsar glitches providing fitting formulas for the glitch magnitudes and rates. In Sec. IV, we calculate the mismatch (relative loss of the squared SNR) that a single glitch will cause. Finally, in Sec. IV, we translate the observed glitches into a prediction for mismatches during a few selected current and future continuous-wave searches and discuss the risk faced by CW searches from glitches.

## II. CONTINUOUS GRAVITATIONAL-WAVE SEARCHES

Searches that target a known pulsar making use of the observed EM emission (for example, the targeted search for the Crab and Vela pulsars by Aasi *et al.* [17]) are able to handle the epoch of a glitch, either by avoiding searches over the glitch or allowing for a jump in the timing solution at that point [18,19]. By this merit, such searches have a very low risk of being disrupted by a glitch coupled to the EM channel, provided the CW channel closely follows the phase evolution of the EM channel in between glitches.

In contrast, wide parameter-space CW searches that, by definition, search for signals without an EM counterpart do not have any such prior knowledge. This category of searches includes both *directed* searches, in which a single sky point in which a neutron star is believed to exist is searched (see, for example, the work by Aasi *et al.* [20], Wette *et al.* [21], and Zhu *et al.* [22]), and *all-sky* searches; in both instances, a band of frequencies and frequency derivatives are usually searched since they are inherently unknown. These searches use a matched filter against smooth templates built from a Taylor expansion in the phase; as such, they do not include glitches. If a neutron star emitting detectable levels of CW emission undergoes a glitch in the CW channel, then the matched-filtering method will not behave as expected because the template is a poor match to the real, glitching signal.

CW searches ideally employ a *fully coherent* search method that consists of matched filtering the template against all the data. However, such a search is typically computationally infeasible, and so CW searches use a *semicoherent* method: the total observation time $T$ is divided into $N_{\text{seg}}$ segments of duration $T_{\text{seg}}$. Each of these segments is fully coherently analyzed and then recombined incoherently to give a semicoherent measurement, which is insensitive to phase jumps between segments. This method provides more sensitive searches at fixed computing cost [23,24]. Typically, a semicoherent search is performed first; then, interesting candidates are *followed up* with longer coherent integration times by reducing the number of segments, aiming to eventually confirm a signal with a final fully coherent search; see the work by Shaltev and Prix [25] for a discussion of a two-stage follow-up procedure and by Papa *et al.* [26] for a multistage application.

## III. STATISTICAL PROPERTIES OF THE OBSERVED GLITCHES

In this section, we study the properties of glitches in the observed radio pulsar population using the glitch catalog maintained by Espinoza *et al.* [27] and available at www.jb.man.ac.uk/pulsar/glitches.html. Our goal is to make a statement about how often glitches occur and their magnitudes for the types of neutron star which may be emitting CWs in the parameter space of typical CW searches. This task is made difficult since many searches look for young, rapidly spinning-down stars for which we only have a small sample of observations or no observations at all. Therefore, we must extrapolate the glitch properties for the population of CW-emitting neutron stars from the observed radio pulsar population.

Radio pulsar timing methods detect glitches by fitting a piecewise Taylor-series expansion in the phase to either side of the event, with a modeled jump in between (see the work by Edwards *et al.* [28] for a detailed discussion). The glitch catalog [27] reports 472 events from 165 isolated pulsars (as of June 27, 2016); for each of these events, a value is reported for the frequency jump $\delta\nu$ and frequency derivative $\delta\dot{\nu}$, if it can be measured. We cross-reference the glitch catalog with the Australia Telescope National Facility (ATNF) [29] pulsar catalog available at www.atnf.csiro.au/people/pulsar/psrcat/ in order to obtain the glitching pulsar's timing properties.

Of the 472 listed glitches, we find 15 with no ATNF cross-reference, one with $\delta\nu < 0$ and four with no measured $\dot{\nu}$ in the ATNF catalog; these pulsars are removed from our data set. Additionally, we find 54 glitches that have either no measured $\delta\dot{\nu}$ or a measured value consistent with zero; these will be included.

### A. Glitch magnitudes

Espinoza *et al.* [27] argued that the glitch catalog contains glitches from two distinct subpopulations of pulsars. There is the main population with $\delta\nu$ magnitudes ranging from $10^{-9}$ to $10^{-5}$ Hz (which we will refer to as the "normal" glitches) and a second, smaller population with larger magnitudes of $\delta\nu$, referred to as "Vela-like" because the pulsars undergoing these glitches have similar characteristic ages and magnetic field strengths to the Vela pulsar (PSR B0833-45). We reproduce the evidence for this finding in Fig. 1 where we plot the histogram of all observed $\delta\nu$ values. This illustrates the bimodality found by Espinoza *et al.* [27].

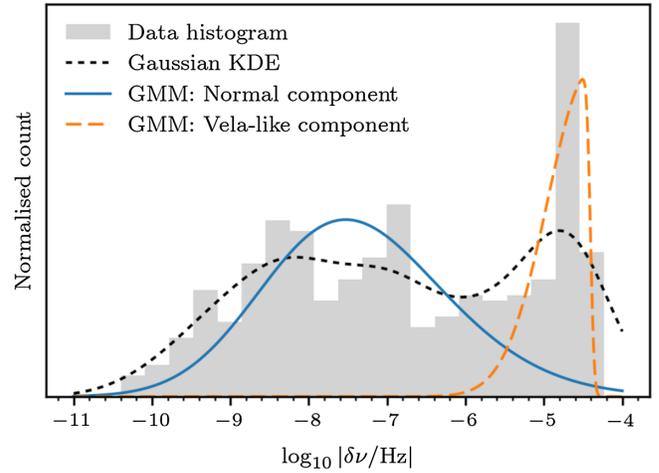

FIG. 1. The distribution of observed glitch magnitudes $\delta\nu$ (from the glitch catalog). This is given as both a binned histogram and a Gaussian KDE, as discussed in the text. The solid and long-dashed lines mark the two components of the two-component skewed Gaussian mixture model (GMM).

To check that the bimodality is not an artifact of the histogram bin sizes, we estimate the probability density function using a Gaussian kernel density estimate (KDE). Specifically, we use the Jones *et al.* [30] implementation. This is also plotted in Fig. 1 and shows two distinct peaks, although the lower peak could also be interpreted as two peaks close together. By eye, it is clear that there are at least two modes to the histogram and possibly more. We investigate this in Appendix A by applying a Bayesian model comparison for Gaussian mixture models (see the work by Gelman *et al.* [31] for a review), varying the number of components, and also allowing for a skew as described by O'Hagan and Leonard [32]. We find that all models with two or more components fit the data decisively better than a single component. Marginal gains are found by allowing the models to be skewed and have four or more components, but no single model is outstanding among the others. For this reason, we choose to use a two-component model with skew; this provides a good empirical description of the data and is pragmatic in that we limit the number of components to two for interpretability. We note that this description is empirical, and we do not intend to make any substantive claim regarding the underlying physics of the two components.

The mixture components and individual distributions for the two-component skewed model are plotted in Fig. 1, and in Table I we provide the resulting mean, standard deviation, weights, and skewness of the two components in log space. This method identifies the two subpopulations in a manner consistent with the observations by Espinoza *et al.* [27], and notably the Vela-like component suffers a significant skew.

TABLE I. The properties of the fitted two-component skewed Gaussian mixture model shown in Fig. 1.

|          | Mean   | Standard deviation | Weight | Skew    |
|----------|--------|--------------------|--------|---------|
| Normal   | −8.391 | 1.591              | 0.700  | 1.056   |
| Vela-like| −4.406 | 0.545              | 0.300  | −9.949  |

We use the best-fit two-component skewed Gaussian mixture model to label each data point as originating from one of two skewed Gaussian distributions. Specifically, to each data point, we assign the label based on the maximum probabilities of each of the two components, given the maximum posterior model parameters derived in the fitting process. In Fig. 2, we plot histograms for $\delta\nu$ and $\delta\dot\nu$ along with the raw data in a scatter plot. We have separated the data into the individual subpopulations, as labeled by the two-component skewed Gaussian mixture model. Several pulsars of interest are picked out using contrasting markers. It is interesting that not all of the Vela glitches are categorized by this method as Vela-like, which can be seen by looking at the distribution of Vela glitches in Fig. 2.

### B. Overview of the population of glitches

To give an overview of all observed glitches in the context of the wider population of observed radio pulsars listed in the ATNF catalog, in Fig. 3, we plot two copies of the familiar $\nu$-$\dot\nu$ diagram. In panel (a), for each pulsar that has been observed to glitch, we add a filled circle with an area proportional to the number of glitches seen in that pulsar. In panel (b), we mark each pulsar that has been observed to glitch with a filled circle, but here the area of the filled circles marks the pulsar's average glitch magnitude. For both plots, different shapes have been used to partition the Vela-like and normal glitches (note that some pulsars display glitches from both populations). Finally, dashed lines mark isoclines of constant characteristic age, $\tau_{\rm age} = |\nu/\dot\nu|$.

While the bulk of observed glitching pulsars are from the main pulsar population, the fraction of young pulsars ($\tau_{\rm age} < 10^5$ yr) that glitch is proportionally higher than in the normal population. Vela-like glitches occur predominantly in the young pulsars with none seen in pulsars with $\tau_{\rm age} > 10^7$ yr. It is also noticeable that younger pulsars display a greater number of glitches. Note that, since we have not observed all pulsars for the same duration, one cannot infer the relative glitch rate from the number of glitches alone.

For the normal-glitch population, Espinoza *et al.* [27] noted that "pulsars with $\tau_{\rm age} < 5 \times 10^3$ yr undergo small- or medium-sized glitches ($\delta\nu < 10^{-5}$ Hz)." It is postulated that the higher temperatures in younger pulsars prevent the glitch mechanism from working effectively. This effect is consistent with Fig. 3(b); the pulsars with the largest average glitch sizes have $\tau_{\rm age} \sim 10^5$ yr, while younger pulsars tend to exhibit smaller glitches on average.

### C. Extrapolating: Glitch magnitudes

We would like to be able to predict the glitch magnitude for the unobserved neutron star population targeted by CW searches. In particular, we need to extrapolate up to large spin-down rates $-\dot\nu \sim (10^{-9}$–$10^{-7})$ Hz/s searched for in many recent CW searches, which are larger than what has been observed in radio pulsars.

It has previously been found [27,33–35] that the glitch *activity* (defined in the first of these references) correlates well with $|\dot\nu|$ and the characteristic age $\tau_{\rm age}$. We choose not to combine the rate and magnitude information together into the activity but estimate both separately as these are of most direct relevance to CW searches.

We investigate correlations of the glitch magnitudes $\delta\nu$ and $\delta\dot\nu$ with the frequency $\nu$, frequency derivative $\dot\nu$, and characteristic age $\tau_{\rm age}$, as shown in Table II. This is done for three groups: all the data together and individually for the normal population and the Vela-like population. For the normal population, both glitch magnitudes most strongly correlate with the spin-down rate $\dot\nu$, although we recognize that the correlation with $\tau_{\rm age}$ is only marginally weaker. In contrast, $\delta\nu$ for the Vela-like population has a weak correlation with all predictor variables, but $\delta\dot\nu$ correlates well with $\dot\nu$ and most strongly with the characteristic age. For simplicity, we choose to use $\dot\nu$ as a predictor variable for both the normal and Vela-like populations, making it simpler to interpret later results as the same predictor is used for both populations. In practice, our conclusions will be robust to either choice of predictor variable.

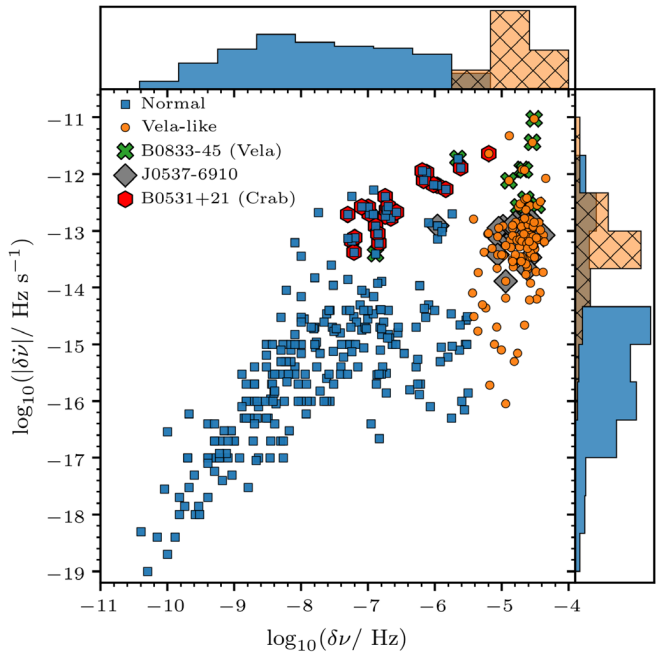

FIG. 2. Scatter plot of all pairs of $\delta\nu$ and $\delta\dot\nu$ in the glitch database: squares indicate the glitches labeled as normal by the mixture model, while circles are those points labeled as Vela-like. Histograms for the glitch magnitudes are also given for each subpopulation; the hatching indicates the Vela-like population histogram. Markers highlight the glitches from interesting pulsars.

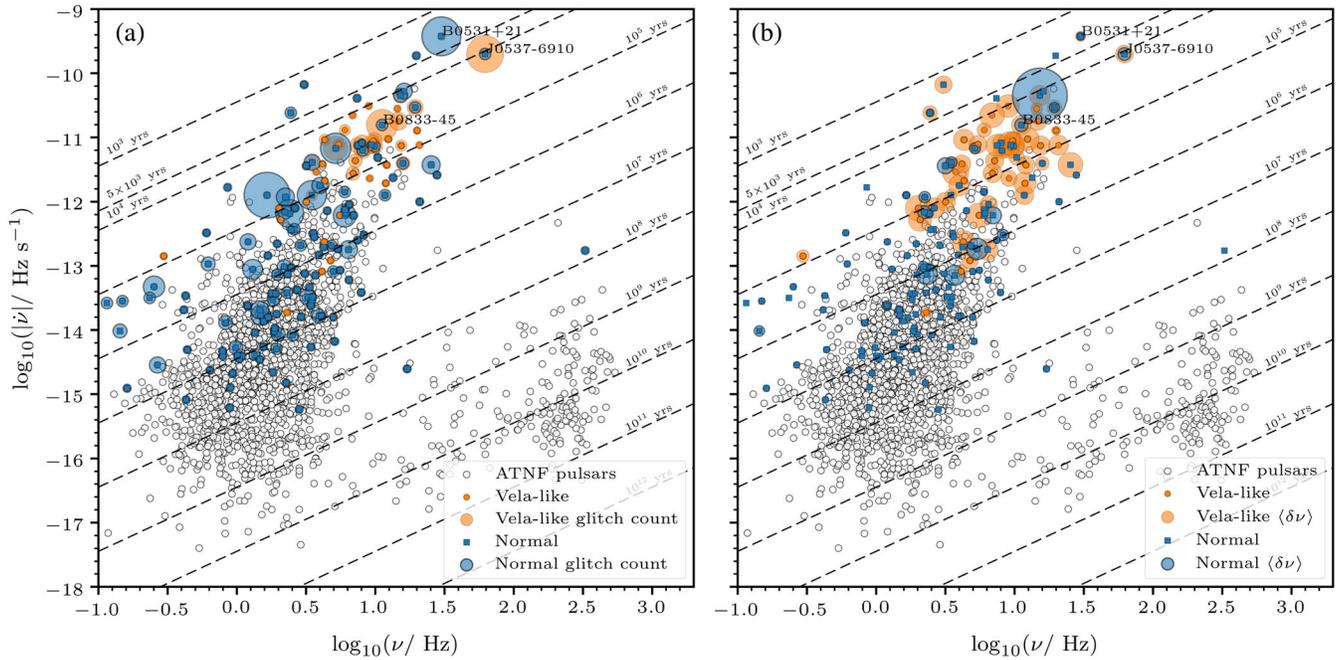

FIG. 3. $\nu$-$\dot{\nu}$ plot of all pulsars in the ATNF catalog [29], overlaid with squares for the normal glitching pulsars and circles for the Vela-like glitching pulsars. (a): For each pulsar observed to glitch, the area of a surrounding filled circle is proportional to the number of observed glitches from that pulsar. (b): For each pulsar observed to glitch, the area of the filled circle is proportional to the average glitch magnitude from that pulsar. Note that in the plot showing glitch magnitudes the relative scaling of the Vela-like and normal populations are *not* equal; the area representing the normal glitch magnitudes are scaled to be three times larger than the Vela-like glitch magnitudes.

In Fig. 4(a) and Fig. 4(b), we show scatter plots of glitch magnitudes against the spin-down rate of the pulsar to demonstrate the correlation. For both plots, we have added contrasting markers to label several interesting pulsars. These help to show that there can be almost as much variation in the glitch magnitude of a single pulsar as from the entire population.

Fitting a linear function in log-log space (see Appendix B for details), our resulting fitting formulas for the frequency jump for each of the two populations is

$$\langle \delta\nu \rangle_{\text{Normal}} = 10^{-0.90} |\dot{\nu}|^{0.55} 10^{\pm 0.93} \quad (1)$$

$$\langle \delta\nu \rangle_{\text{Vela-like}} = 10^{-4.59} |\dot{\nu}|^{0.02} 10^{\pm 0.28} \quad (2)$$

TABLE II. The correlation coefficients between the glitch magnitudes and the timing properties of the source pulsar.

|  |  | $\log_{10} |\tau_{\text{age}}|$ | $\log_{10} |\nu|$ | $\log_{10} |\dot{\nu}|$ |
|---|---|---|---|---|
| All | $\log_{10} |\delta\nu|$ | −0.634 | 0.538 | **0.68** |
|  | $\log_{10} |\delta\dot{\nu}|$ | −0.846 | 0.672 | **0.88** |
| Normal | $\log_{10} |\delta\nu|$ | −0.631 | 0.390 | **0.64** |
|  | $\log_{10} |\delta\dot{\nu}|$ | −0.864 | 0.604 | **0.88** |
| Vela-like | $\log_{10} |\delta\nu|$ | 0.037 | **0.13** | 0.048 |
|  | $\log_{10} |\delta\dot{\nu}|$ | **−0.62** | 0.376 | 0.593 |

and for the frequency-derivative jumps is

$$\langle \delta\dot{\nu} \rangle_{\text{Normal}} = 10^{-4.17} |\dot{\nu}|^{0.90} 10^{\pm 0.67} \quad (3)$$

$$\langle \delta\dot{\nu} \rangle_{\text{Vela-like}} = 10^{-7.03} |\dot{\nu}|^{0.57} 10^{\pm 0.66}, \quad (4)$$

where the last factor provides an estimate of the variability about the linear fit. These fits do not provide a precise statement about the magnitude of glitches but are sufficient to estimate the orders of magnitude that we might expect.

Our data set includes one glitch from a millisecond pulsar, PSR B1821-24 [36]; it is interesting to note that the glitch magnitude is typical of normal pulsars with similar spin-down rates. This is also true for the second observation of a glitch in a millisecond pulsar, PSR J0613-0200 [37] (not included in the data set as it was added to the glitch database after the analysis was completed). This indicates that, despite the much greater spin frequency of these millisecond pulsars, the glitch magnitude depends on the spin-down rate in much the same way as seen for normal pulsars.

### D. Extrapolating: Average glitch rate

To estimate the average rate of glitches, Espinoza *et al.* [27] grouped pulsars by their spin-down rate $\dot{\nu}$, including pulsars that have not yet been observed to glitch. From this grouping, the authors used the measured number of glitches $N_g$ to calculate a mean glitch rate $\langle \dot{N}_g \rangle$. In Fig. 10 of their work, they show that, to a good approximation, in log space, the mean glitch rate depends linearly on the spin-down rate;

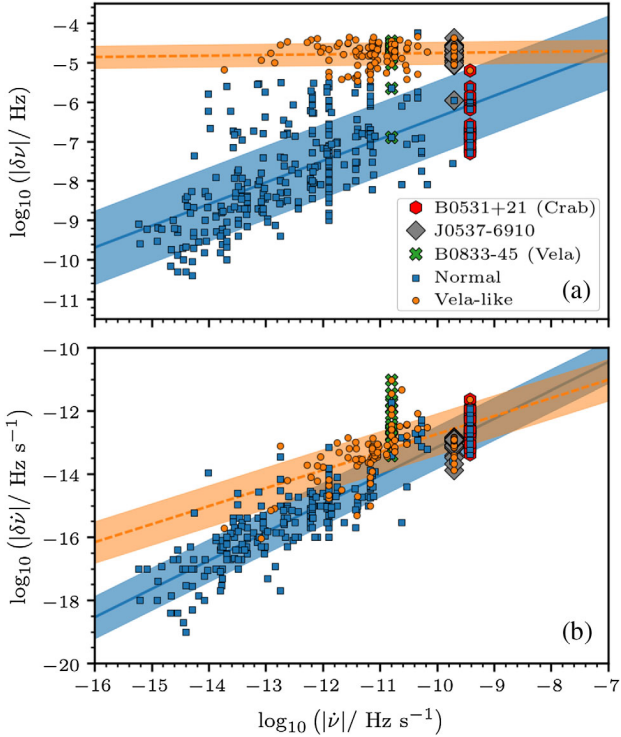

FIG. 4. (a): The magnitude of $\delta\nu$ for the normal glitches [solid line, given by Eq. (1)] and Vela-like glitches [dashed line, given by Eq. (2)] as a function of the pulsar's spin-down rate. (b): The magnitude of $\delta\dot{\nu}$ for the normal glitches [solid line, given by Eq. (3)] and Vela-like glitches [dashed line, given by Eq. (4)] as a function of the measured spin-down rate. The shaded region marks the variability about the linear fit. Vertical clustering in the observed data points is the result of multiple glitches observed from a single source. Markers highlight glitches from some interesting pulsars.

we reproduce this in Fig. 5 using the data from Table 4 by Espinoza et al. [27].

To extrapolate, we fit a linear function to the glitch rate and find the fitting formula

$$\langle \dot{N}_g \rangle = 10^{-3.00} |\dot{\nu}|^{0.47} 10^{\pm 0.31} \text{ s}^{-1}, \quad (5)$$

where $\dot{\nu}$ is measured in Hz/s. The exponent agrees with that found by the original authors (they do not provide the prefactor).

## IV. SIGNAL LOSS DUE TO GLITCHES

After the discussion in the previous section of the characteristics of glitches seen in radio pulsars, we will now answer the question of what the effect of a glitch in a CW signal on matched-filtered searches is. In the next section, this will be combined with the predictions of the previous section to try and quantify the risk posed by glitches to CW searches. We begin this section by introducing the mismatch, followed by analytic and numerical estimates for the mismatch of glitching signals.

### A. General mismatch definitions

We assume the template family consists of a regular CW phase model (without glitches) with phase parameters $\theta$, defined at some reference time $t_{\text{ref}}$:

$$\Phi_t(t) \equiv \Phi(t - t_{\text{ref}}; \theta). \quad (6)$$

The mismatch $\mu^{(0)}$ is generally defined[2] as the relative loss of the (squared) signal-to-noise ratio $\rho^2$ due to an offset between signal parameters $\theta_s$ and template parameters $\theta$, namely,

$$\mu^{(0)}(\theta_s; \theta) \equiv \frac{\rho^2(\theta_s; \theta_s) - \rho^2(\theta_s; \theta)}{\rho^2(\theta_s; \theta_s)} \in [0, 1], \quad (7)$$

where $\rho^2(\theta_s; \theta)$ denotes the expected squared SNR in template $\theta$ in the presence of a signal[3] with parameters $\theta_s$ and $\rho_s \equiv \rho(\theta_s; \theta_s)$ is the maximal "perfect-match" SNR. When considering the loss of the SNR due to a glitch (or timing noise) in the emitting neutron star, we are in the situation of signals that lie *outside* of the regular CW template manifold $\mathbb{T}$, as originally discussed by Apostolatos [39] and Owen [40]. This effect can be quantified by the "fitting factor" $FF$ [39] or equivalently the *minimal mismatch* $\mu_{\min}^{(0)}$ in the limit of an infinitely finely spaced template bank, i.e.,

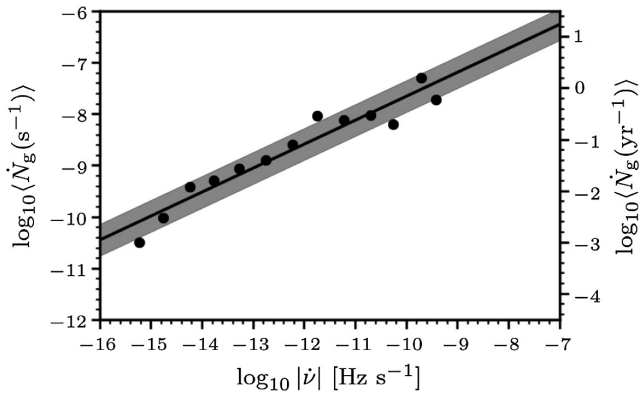

FIG. 5. Reproduction of Fig. 10 from Espinoza et al. [27], giving both the log of the glitch rate per second (left-hand axis) and per year (right-hand axis). Black dots are the original data points, while the solid line and shaded region are our best-fit line and a measure of the variability as given in Eq. (5).

---

[2]Obviously, using any quantity that is merely *proportional* to the SNR would also yield the same mismatch definition.
[3]For simplicity in the following, we are suppressing the dependency on polarization parameters; see the work by Prix [38] for more detailed discussion of this common "phase metric" approximation.

$$\mu_{\min}^{(0)}(\theta_s) \equiv \min_{\theta \in \mathbb{T}} \mu^{(0)}(\theta_s; \theta). \quad (8)$$

Evidently, if the signal lies inside the template manifold $\mathbb{T}$, i.e., if there is a $\theta = \theta_s \in \mathbb{T}$, then the minimal mismatch would be zero, while for signals outside of the template manifold, it would be nonzero. Therefore, any additional mismatch from a finitely spaced template bank will further increase the total loss of the SNR, but in the following, we are only interested in the minimal match for glitching signals, which is independent of template-bank spacing.

The *metric mismatch* $\mu$ is a common and useful approximation [40–42] to the (full) mismatch $\mu^{(0)}$ of Eq. (7) for small offsets $\Delta\theta \equiv \theta - \theta_s$ between template and signal; i.e., by Taylor expanding, we obtain

$$\mu^{(0)}(\theta_s; \theta) = \mu(\theta_s; \theta) + \mathcal{O}(\Delta\theta^3), \quad \text{with} \quad (9)$$

$$\mu(\theta_s; \theta) \equiv g_{ij}(\theta_s)\Delta\theta^i \Delta\theta^j \in [0, \infty), \quad (10)$$

where $g_{ij}$ is referred to as the *metric*.

The metric mismatch of Eq. (10) is generally a good approximation for small mismatches, and empirically, one finds that deviations between $\mu^{(0)}$ and $\mu$ start to become noticeable for mismatches above $\mu \gtrsim 0.3$–0.5 (e.g., see Refs. [38,43]). For larger offsets, the (unbounded) metric mismatch will tend to increasingly overestimate the actual (bounded) mismatch $\mu^{(0)}$, which only slowly asymptotes to 1, corresponding to a total loss of the SNR. See the work by Wette [44] for a more detailed discussion of this effect and an empirical fit to extend the applicability of the metric mismatch. This distinction will be relevant in the following discussion, as glitches can result in large parameter-space offsets $\Delta\theta$ compared to the range of validity of the metric approximation, and therefore one needs to be careful in interpreting results obtained via the metric mismatch.

### B. Glitching CW signals

The model of a signal containing $N_g$ glitches[4] is defined as a piecewise CW phase function of the form of Eq. (6) over $N_g + 1$ continuous "domains" labeled by $a, b, \ldots \in [0, N_g]$; i.e., $a = 1$ is the domain following the first glitch, etc. For each continuous stretch $a$, the signal phase function is therefore

$$\Phi_s(t) = \Phi(t - t_{\text{ref}}; \theta_{(a)}), \quad \text{for } t \in [t_{(a)}, t_{(a+1)}), \quad (11)$$

and the length of each continuous stretch $a$ is $T_{(a)} \equiv t_{(a+1)} - t_{(a)}$. We further denote the fractional length of each domain as

$$R_{(a)} \equiv \frac{T_{(a)}}{T}. \quad (12)$$

A neutron-star glitch would produce a jump $\delta' f$ in signal frequency and a jump $\delta' \dot{f}$ in the spin-down. We will also consider a phase jump $\delta' \phi$. Such a jump in the phase is not normally considered in radio pulsar astronomy, as the uncertainty in the time $t_{\text{ref}}$ at which a glitch occurs is much greater than the pulsar spin period. This is related to the fact that glitches often contain initial transient jump components in frequency and spin-down that decay away on various time scales (hours to days or longer) [34]. Given that these transient glitch components are not included in our simplified glitch model, we can effectively consider the *asymptotic* persistent jump to also contain a phase jump $\delta' \phi$. Such phase jumps are potentially relevant to gravitational-wave searches, however, and so need to be included in our analysis.

We therefore define glitch $a$ as a discontinuity at time $t_{(a)}$ in signal parameters with offset $\{\delta' \phi_{(a)}, \delta' f_{(a)}, \delta' \dot{f}_{(a)}\}$, while all other signal parameters (e.g., sky position $\vec{n}$ and higher-order spin-down terms $\ddot{f}, \ldots$ and binary orbital parameters) remain constant. We denote the phase parameters explicitly as $\theta = \{\phi, f, \dot{f}, \ddot{f}, \ldots\}$, so the jump induced by glitch $a$ would be $\delta' \theta_{(a)} = \{\delta' \phi_{(a)}, \delta' f_{(a)}, \delta' \dot{f}_{(a)}, 0, 0, \ldots\}$. Successive jumps would be cumulative, and so we define

$$\delta\theta_{(a)} \equiv \sum_{b=1}^{a} \delta' \theta_{(b)}, \quad \text{and} \quad (13)$$

$$\theta_{(a)} = \theta_{(0)} + \delta\theta_{(a)} \quad \text{for } a \geq 1, \quad (14)$$

where the cumulative effect is a simple sum because all phase parameters $\theta$ refer to the same fixed reference time $t_{\text{ref}}$ and are therefore constant in time (in between glitches). The offset between signal and template parameters in each stretch $a$ is denoted as

$$\Delta\theta_{(a)} \equiv \theta_{(a)} - \theta. \quad (15)$$

### C. Single-glitch metric mismatches

The general expressions for the metric mismatch for $N_g$ glitches are derived in Appendix C. Here, we only present the explicit results obtained for the special case of a single glitch, occurring after time $T_{(0)} = RT$, where $T$ is the total observation time and where we average over the unknown fraction $R \in [0, 1]$. These metric results are limited to the case of "directed searches" in which the SNR is maximized over a search space $\{f, \dot{f} \ldots\}$, but the sky position of the source is assumed to be known.

#### 1. Coherent searches

Consider three different search scenarios: a four-dimensional (4D) search covering $\{f, \dot{f}, \ddot{f}, \dddot{f}\}$, a 3D search including up to $\dddot{f}$, and a 2D search covering only $\{f, \dot{f}\}$. For these three cases, we explicitly (see Appendix C 3) find the $R$-averaged minimal coherent glitch mismatches as

$$\tilde{\mu}_{\min}^{4D} = \frac{5\delta\phi^2}{198} + \frac{5\pi T^2}{18018}(\pi\delta f^2 - \delta\phi\delta\dot{f}) + \frac{\pi^2 T^4}{540540}\delta\dot{f}^2, \quad (16)$$

$$\tilde{\mu}_{\min}^{3D} = \frac{2\delta\phi^2}{63} + \frac{2\pi T^2}{3465}(\pi\delta f^2 - \delta\phi\delta\dot{f}) + \frac{\pi^2 T^4 \delta\dot{f}^2}{135135}, \quad (17)$$

$$\tilde{\mu}_{\min}^{2D} = \frac{3\delta\phi^2}{70} + \frac{\pi T^2}{630}(\pi\delta f^2 - \delta\phi\delta\dot{f}) + \frac{\pi^2 T^4}{13860}\delta\dot{f}^2. \quad (18)$$

---
[4]Here and in the following, we consider a general sequence of glitches, which would also allow one to model the effect of timing noise, considered as a sequence of small "glitches" (and allowing for both positive and negative jumps in frequency). However, in order to simplify the terminology, we continue to refer to the effect as glitches.

These mismatches are minimal in the sense defined by Eq. (8), i.e., the mismatches that remain once one has searched over an infinitely finely spaced template bank. These expressions illustrate the effect of correlations in parameter space: including more search dimensions lowers the minimal achievable glitch mismatch over the template bank. In particular, including $\ddot{f}$ as a search dimension compared to only $\dot{f}$ reduces the mismatch of a pure spin-down jump $\delta\dot{f}$ by a factor of $\sim 10$, a pure frequency jump $\delta f$ by a factor of $\sim 3$, and a pure phase jump $\delta\phi$ by roughly $\sim 35\%$. Including both $\ddot{f}$ and $\dddot{f}$ reduces the mismatch even more: a spin-down jump by a factor of 39, a frequency jump by a factor of $\sim 6$, and a phase jump by $\sim 70\%$ compared to a search up to $\dot{f}$. Similarly, including sky position $\vec{n}$ as a search dimension would likely further reduce the achievable minimal mismatch as well, we investigate this in Sec. IV E 2, but there is no simple analytic way to estimate this effect due to the intrinsic complications of the sky metric [43].

Note, however, that as a consequence of this "compensation" effect there will be biases in the estimated parameters of a glitching signal, which can lead to problems when following up or interpreting such candidates without taking into account the fact that the signal has potentially undergone a glitch.

As an interesting application of these $R$-averaged metric expressions, we can derive maximal coherent observation times $T$ for a given glitch jump in either $f$ or $\dot{f}$ (using canonical values corresponding to the largest observed jumps) and a maximal tolerated glitch mismatch (using a canonical value of 0.1). For pure frequency jumps $\delta f$, we find

$$T \leq \{7.0, 4.8, 2.9\} \text{ days} \left(\frac{\tilde{\mu}}{0.1}\right)^{1/2} \left(\frac{|\delta f|}{10^{-5} \text{ Hz}}\right)^{-1}, \quad (19)$$

for a 4D, 3D, or 2D search, respectively. In the case of a pure spin-down jump $\delta\dot{f}$, we find

$$T \leq \{100, 70, 40\} \text{ days} \left(\frac{\tilde{\mu}}{0.1}\right)^{1/4} \left(\frac{|\delta\dot{f}|}{10^{-12} \text{ Hz/s}}\right)^{-1/2}. \quad (20)$$

### 2. Semicoherent searches

In a semicoherent search, the total observation time $T$ is divided into $N_{\text{seg}}$ segments of length $T_{\text{seg}} = T/N_{\text{seg}}$, each segment $\ell$ is analyzed coherently, and the results are combined incoherently by summing the "power" from each segment (e.g., see the work by Prix and Shaltev [23] and Brady and Creighton [24]).

For simplicity, we assume the glitch to happen on a segment boundary, after $N_{(0)} = RN_{\text{seg}}$ segments. If a glitch happens inside a segment, the additional mismatch in that segment is bounded within [0, 1] and at most contributes $1/N_{\text{seg}}$ to the total mismatch.

From the detailed derivations in Sec. C 4, we can obtain explicit $R$-averaged glitch mismatch expressions in the limit of a large number of segments $N_{\text{seg}}$ in the three different search spaces (4D = $\{f, \dot{f}, \ddot{f}, \dddot{f}\}$, 3D = $\{f, \dot{f}, \ddot{f}\}$ and 2D = $\{f, \dot{f}\}$), namely,

$$\hat{\mu}_{\min}^{4D} = \frac{2\pi^2 T_{\text{seg}}^2}{189}\delta f^2 + \frac{\pi^2 N_{\text{seg}}^2 T_{\text{seg}}^4}{20790}\delta\dot{f}^2, \quad (21)$$

$$\hat{\mu}_{\min}^{3D} = \frac{\pi^2 T_{\text{seg}}^2}{70}\delta f^2 + \frac{\pi^2 N_{\text{seg}}^2 T_{\text{seg}}^4}{7560}\delta\dot{f}^2, \quad (22)$$

$$\hat{\mu}_{\min}^{2D} = \frac{\pi^2 T_{\text{seg}}^2}{45}\delta f^2 + \frac{\pi^2 N_{\text{seg}}^2 T_{\text{seg}}^4}{1260}\delta\dot{f}^2. \quad (23)$$

Here, we see again how parameter-space correlations act to reduce the minimal-glitch mismatch when including more search dimensions. Note that the phase jump $\delta\phi$ does not appear in these expressions, as in a semicoherent search the SNR in each segment is separately maximized over the phase of the signal within that segment, so that phase jumps in between segments have no effect.

Turning these $R$-averaged expressions into bounds on the coherent segment length for a pure frequency jump, we find

$$T_{\text{seg}} \leq \{1.1, 1.0, 0.8\} \text{ days} \left(\frac{\tilde{\mu}}{0.1}\right)^{1/2} \left(\frac{\delta f}{10^{-5} \text{ Hz}}\right)^{-1}, \quad (24)$$

while for a pure spin-down jump, we obtain

$$T_{\text{seg}} \leq \{5.3, 3.2, 1.3\} \text{ days}$$
$$\times \left(\frac{\tilde{\mu}}{0.1}\right)^{1/2} \left(\frac{T}{365 \text{ days}}\right)^{-1} \left(\frac{\delta\dot{f}}{10^{-12} \text{ Hz/s}}\right)^{-1}. \quad (25)$$

### D. Upper bounds on glitch mismatches

#### 1. Coherent searches

The metric mismatches discussed in the previous section can be a useful tool, for example, to estimate bounds on the maximal safe observation span in the presence of a glitch, as given in Eqs. (16)–(18) and Eqs. (21)–(23). However, the metric is of somewhat limited use for predicting mismatches from glitches in general because the parameter jumps $\delta\theta$ will often not be small enough for the metric approximation of Eq. (C12) to be applicable. In fact, it is the regions of "large" mismatches $\sim \mathcal{O}(1)$ that would be most relevant for a realistic assessment of the impact of glitches on the detectability of signals. While numerical extrapolations to the metric mismatch do exist for large template mismatches [44], it is unclear to what extent they would be applicable in the present case of off-manifold signal mismatches. However, we can also obtain simple analytic upper bounds on the mismatches, which can be useful to estimate the relevant "scale" of the expected mismatches.

Namely, we can always find a template that perfectly matches the signal over the *longest* time stretch $a'$, i.e., $R_{(a')} = \max_a R_{(a)}$, during which the difference between the template and signal is zero. While it is in principle possible for other time stretches $a \neq a'$ to *reduce* the total SNR by noticeable negative contributions in the (complex) coherent matched-filtering amplitude [cf. Eq. (C9)], this is generally quite unlikely: offsets in frequency $\Delta f_{(a)}$ or spin-downs $\Delta \dot{f}_{(a)}$ would result in a rapidly oscillatory matched-filtering integrand $e^{i\Delta\Phi(t)}$ [where $\Delta\Phi(t)$ is the difference between the signal phase and the template phase defined in Eq. (C8)], and so the corresponding contributions would typically be small, while only pure phase jumps $\Delta\phi_{(a)} \sim \pi$ could result in canceling contributions. Therefore, generally, the matched-filtering amplitude of Eq. (C11) will be $X = \max_a R_{(a)}$ (assuming the template is perfectly matched in this domain but contributes neither positively nor negatively elsewhere). And so we obtain an upper bound on the coherent glitch mismatch as

$$\tilde{\mu}_{\min}^{(0)} \lesssim 1 - (\max_a R_{(a)})^2 \leq 1 - \frac{1}{(N_g + 1)^2}, \quad (26)$$

where the second inequality is obtained by observing that $\max_a R_{(a)} \geq \frac{1}{N_g+1}$.

### 2. Semicoherent searches

By regrouping Eq. (C39) over the interglitch time stretches, we can write the semicoherent glitch mismatch as

$$\hat{\mu}^{(0)} = \sum_{a=0}^{N_g} R_{(a)} \hat{\mu}_{(a)}^{(0)}. \quad (27)$$

Using the same argument as before, there will always be a template that perfectly matches the signal over the longest interglitch time stretch $\max_a R_{(a)} T$, where the mismatch would be zero, while it would be at most 1 the rest of time (but quite possibly less). This yields the (strict) upper glitch mismatch bound of

$$\hat{\mu}_{\min}^{(0)} \leq 1 - \max_a R_{(a)} \leq \frac{N_g}{N_g + 1}, \quad (28)$$

which, contrary to the coherent case of Eq. (26) is *linear* in $\max_a R_{(a)}$ instead of quadratic.

### E. Numerical estimates of the mismatch

The metric estimates discussed in Sec. IV C describe the behavior of small glitch mismatches, while the results of Sec. IV D provide useful analytic upper bounds. To verify these expressions as well as interpolate between the two regimes of small and maximal glitch mismatches, we also use direct numerical computation of the mismatch. In this section, we present two different approaches to numerically estimate the mismatch: (i) a relatively simpler numerical evaluation of the simplified matched-filter amplitude of Appendix C 1, with a minimization over the template parameters, and (ii) direct evaluation of the full CW $\mathcal{F}$ statistic [45] over the template space using an efficient Markov-chain Monte Carlo (MCMC) search.

Understanding the intermediate regime between metric estimates and upper mismatch bounds will be important when quantifying the risk posed by glitches to real searches, as glitches will often be sufficiently large to cause substantial mismatches ($\gtrsim 0.3$), as was seen already from the simple estimates in Eqs. (19)–(20) and Eqs. (24)–(25).

#### 1. Mismatch from simplified matched-filter amplitude

Given a particular glitching signal, we can estimate the mismatch by numerically evaluating the simplified matched-filtering amplitude $X$ defined in Eq. (C8) over the template manifold. This approach therefore only involves the differences between the template phase model and the signal phase model, while still neglecting antenna-pattern and polarization parameters.

Calculating the fully coherent mismatch involves generating two time series, namely, the signal- and template- phase functions. The phase difference $\Delta\Phi(t) = \Phi_s(t) - \Phi_t(t)$ is then used to compute the simplified matched-filtering amplitude $X$, and the fully coherent mismatch is estimated as $1 - |X|^2$, as discussed in Appendix C 1, and numerically minimized over the template search parameters using a Nelder-Mead minimization (the Jones *et al.* [30] implementation), which was found to be effective for small mismatches. For larger mismatches, however, the mismatch topology contains multiple minima, and this method requires careful selection of the initial guess.

In Fig. 6, we show the results of a simple Monte Carlo study using this approach in the context of a semicoherent search. For each simulation, we generate a signal lasting 100 days with a single fixed glitch of magnitude $\delta f = 5 \times 10^{-7}$ Hz, and we choose the time at which the glitch occurs uniformly over the entire data span. In the upper plot, we show the average minimal mismatch of a 2D search over $\{f, \dot{f}\}$ as a function of the coherent segment length $T_{\text{seg}}$ of the search (and hence the number of segments), comparing the result to the metric prediction of Eq. (23). The fractional difference between the two estimates is shown in the lower plot. For short segment lengths, the mismatch is small $\langle \hat{\mu} \rangle \lesssim 0.1$, and so the numerical estimates agree well with the metric mismatch approximation. However, as the segment length increases, the mismatch grows, and the approximation starts to overestimate the numerical results.

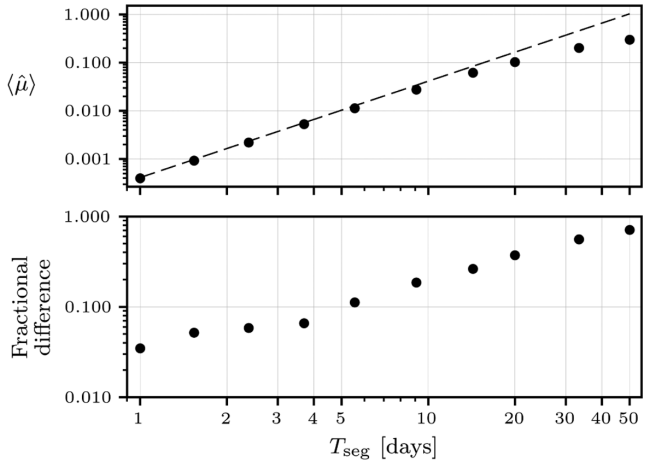

FIG. 6. Comparison of the mismatch (averaged over the time at which a single glitch occurs) of the numerical simplified matched-filtering amplitude of Sec. IV E 1 (solid dots) with the metric mismatch estimate of Eq. (23) (dashed line). For this comparison, an observation time of $T = 100$ days was used, and the glitch had a fixed magnitude of $\delta f = 5 \times 10^{-7}$ Hz.

In addition to verifying the behavior of the metric mismatch approximation, Fig. 6 also demonstrates that for a fairly typical glitch size (compare with Fig. 2, for example) 10% of the squared SNR can be lost for longer segments. Moreover, this also provides insight into what will happen to a signal during a follow-up procedure when the segment length is increased to test the significance of candidates. For short segment lengths, only a small amount of the SNR is lost, but this mismatch will increase as the segment length is increased. Therefore, the SNR will not increase as it is expected to for a CW signal, and the candidate might potentially be classified as "not following the presumed signal model."

### 2. Full $\mathcal{F}$-statistic mismatch using a MCMC search

The second approach consists in directly computing the mismatch of Eq. (7) by evaluating the $\mathcal{F}$ statistic over the template manifold using a MCMC search. For this, we generate data containing a glitching signal (using lalapps_Makefakedata_v5 [46]) and then search for it over the template manifold (not including glitches) using the LALSuite [46] implementation of the $\mathcal{F}$ statistic. The mismatch is then obtained as the fractional difference between the recovered maximal $\mathcal{F}$-statistic value and a perfectly matched signal (including the glitches).

This method allows one to include the effect of a sky search (that is, to answer the question of whether searching over the sky position can further reduce the minimal-glitch mismatch similar to what was found when searching over higher-order spin-downs).

The $\mathcal{F}$ statistic is the log-likelihood ratio of the signal vs Gaussian noise, analytically maximized over the four amplitude parameters $h_0$, $\cos \iota \in [-1, 1]$, $\phi \in [0, 2\pi]$, and $\psi \in [-\pi/4, \pi/4]$, where $\iota$ is the inclination angle of the source, $\phi$ is the initial phase, and $\psi$ is the polarization angle. We compute the mismatch averaged over these amplitude parameters by randomly drawing these amplitude parameters from their prior ranges. Note that we fix the amplitude of the signal to $h_0 = 1$ and generate data without noise in order to obtain the expected $\mathcal{F}$ statistic, which is directly related to the SNR $\tilde{\rho}$ via $E[2\mathcal{F}] = 4 + \tilde{\rho}^2$ [45], which enters the mismatch expression Eq. (7).

To minimize the glitch mismatch, we use a MCMC minimization step with priors for each of the search parameters chosen in order to allow all correlations to be fully explored. A gridded search would yield equivalent results, provided the grid points were sufficiently dense such that the template-bank mismatch was negligible; a MCMC approach (which must equivalently be run for a long enough period to ensure a good approximation of the global maximum) was found to be simpler and computationally less demanding than a gridded search.

As a simple demonstration of this method, in Fig. 7, we show the results of a Monte Carlo study of a semicoherent search (with $T = 60$ days and $T_{\text{seg}} = 15$ days) for a signal containing a single glitch which occurs exactly one-quarter of the way through the search; this setup is chosen such that in the large-glitch limit the maximum mismatch will be exactly 0.25 where the template matches three-fourths segments (see the discussion in Sec. IV D for more details). We plot the mismatch (averaged over amplitude parameters) vs a realistic range of the jump size $\delta f$ in frequency (e.g., see Fig. 2).

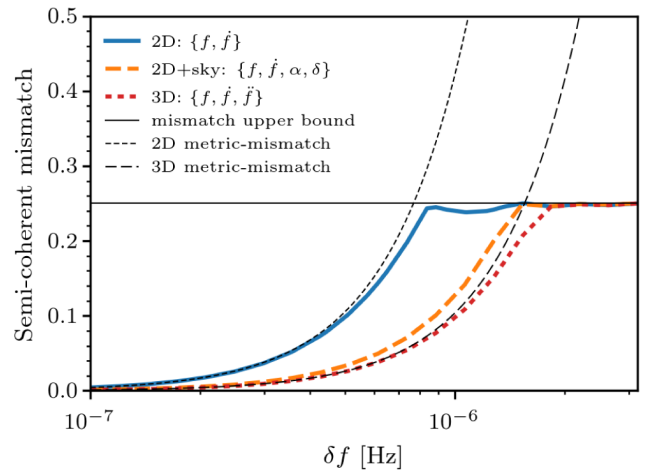

FIG. 7. Semicoherent mismatch as a function of glitch frequency jump $\delta f$ for three different parameter spaces searched (2D=$\{f, \dot{f}\}$, 3D=$\{f, \dot{f}, \ddot{f}\}$, and "2D + sky"). In this example, the search setup is fixed ($T = 60$ days and $T_{\text{seg}} = 15$ days), as is the time of the glitch, which occurs 1/4 of the way through the observation. Thick lines denote the mismatch computed using the $\mathcal{F}$-statistic method of Sec. IV E 2. The horizontal line marks the mismatch upper bound at 0.25, since $\max_a R_{(a)} = 0.75$. Black dashed lines indicate the metric mismatch estimates of Appendix C 3 for the 2D and 3D cases.

This search is repeated for three different parameter spaces: a 2D search over $\{f, \dot{f}\}$, a $(2 + 2)$D search over $\{f, \dot{f}, \text{sky}\}$, and a 3D search over $\{f, \dot{f}, \ddot{f}\}$. In the case of searching over the sky, we found that the minimum mismatches were found up to $\sim 0.1$ rad away from the true sky position, which indicates that glitches can induce substantial biases in the recovered parameters of a search over a nonglitch template bank. The plot also shows the metric mismatch predictions for comparison, as derived in Appendix C 4, without taking the large-$N_{\text{seg}}$ limit (the exact expressions are lengthy and not generally useful and have therefore been omitted).

This confirms and extends the findings of Sec. IV C that searching over more parameters can yield a reduction in mismatch, which also applies to searching over the sky. The glitch mismatch reduction due to minimization over the sky is found to be somewhat smaller but comparable to searching over an extra spin-down parameter $\ddot{f}$.

While the explicit $\mathcal{F}$-statistic search requires greater computational resources than the simplified matched-filtering approach of the previous section, it models the effects of realistic searches most fully and will therefore be used as the method of choice to estimate the risk to real CW searches in the next section.

## V. RISK POSED BY GLITCHES TO CW SEARCHES

In this paper so far, we have provided an empirical study of glitches in the radio pulsar population (Sec. III) and a discussion on how one or more glitches in the CW signal will produce a nonzero mismatch in a matched-filtered gravitational-wave search (Sec. IV). We will now combine these two analyses to assess the possible impact of glitches on some past and possible future CW searches.

Our crucial assumption is that the statistical properties of glitches in the known pulsar population are a good indicator of those of the target population of gravitational-wave emitters. To be more precise, the fitting formulas of Sec. III, for glitch size and rate which used the spin-down rate $\dot{\nu}$ as the indicator variable. For the results to be presented here, assessing the possible impact on CW searches, we will need to extrapolate up to spin-down rates of the order of $10^{-7}$ Hz/s. This is about 2 orders of magnitude larger than the largest spin-down rates seen in the radio pulsars; see Figs. 4 and 5.

Given that the physical mechanism(s) underlying glitches are not well understood, it is difficult to assess the safeness of this assumption, although it seems reasonable to suppose that, whatever the mechanisms are, they will apply to all spinning-down neutron stars (but see the discussion of gravitars, below). Of course, the eventual observation and statistical characterization of glitches in a population of such rapidly spinning-down CW sources would provide a test of this assumption. Note that for spin-down rates with $|\dot{\nu}| > 10^{-7}$ Hz, the fitting formulas for the glitch sizes $\delta \nu$ for the normal and Vela-like populations cross, casting added doubt over their validity at such high spin-down rates; see Fig. 4. The same is true for the $\delta \dot{\nu}$ jumps. Fortunately, we will not need to extrapolate into this regime.

Gravitational-wave astronomers sometimes postulate the existence of a population of *gravitars*, electromagnetically unseen neutron stars that spin down mainly through gravitational-wave emission [47,48]. Such stars would necessarily have very low external magnetic fields, or else they would be visible as pulsars and would be acted upon by significant electromagnetic spin-down torques.

If the internal magnetic field is also small, one might wonder if the glitch mechanism might somehow be suppressed, as compared to the radio pulsar population. We can simply note that for glitches due to superfluid unpinning, if the pinning takes place in the inner crust, the large-scale stellar magnetic fields probably play no significant role [2], so such glitches are still to be expected in gravitars. On the other hand, if, as has been suggested relatively recently [49], pinning takes place in the core, on magnetic flux tubes, the lack of a significant internal magnetic field might indeed suppress this glitch mechanism. For crust quakes, which can be expected to occur alongside whatever superfluid glitch mechanism might be operative, the magnetic field is not expected to play any significant role [4], implying that crust quakes should occur also in gravitars.

To sum up, we can reasonably expect the glitch mechanisms that apply to the radio pulsar population to apply in whole or (at least) in part to the hypothetical gravitar population.

### A. Glitch rate and associated probability

CW searches typically use stretches of data from tens of days to a few months with a small number of searches spanning longer than a year or two. The first question that must be answered to understand the risk is as follows: given that a signal does exist in the data, how probable is it that one or more glitches will occur during the data span?

In Eq. (5), we reproduced the glitch-rate fitting formula for $\langle \dot{N}_g \rangle$ given by Espinoza *et al.* [27], which provides an estimate of the glitch rate per second as a function of the source pulsar's spin-down rate and includes the effect of pulsars that have not been observed to glitch. From this rate, we can estimate the expected number of glitches given a span of data as a function of the spin-down rate. For the glitch magnitude fitting formulas, we split the population of glitches into two subpopulations (normal and Vela-like) to avoid overestimating the glitch magnitude for large spin-down rates. To use these fitting formulas, we will present results in this section similarly split by subpopulation. However, the fitting formulas for the glitch rate were calculated from the whole population; therefore we define

$$\langle N_g \rangle_{\text{normal}}(\dot{\nu}, T) = w_{\text{normal}} \langle \dot{N}_g \rangle T \tag{29}$$

$$\langle N_g \rangle_{\text{Vela-like}}(\dot{\nu}, T) = w_{\text{Vela-like}} \langle \dot{N}_g \rangle T \tag{30}$$

as the *expected number* of normal and Vela-like glitches in which $w_{\text{normal}}$ and $w_{\text{Vela-like}}$ are the weights of the two populations as given in Table I. Implicit in this definition is a prior specification that the proportion of normal and Vela-like pulsars in the target population is the same as in the observed population. There is some evidence that, in fact, the proportion of Vela-like pulsars increases with $\dot{\nu}$; this could be modeled by a $\dot{\nu}$-dependent weighting, but we will ignore this effect here.

The average number of glitches in a given search is a useful quantity, but it is not easy to interpret; for a low average number of glitches, there remains a significant probability of having zero glitches and hence no loss of signal. To better understand this risk, we will therefore apply a simple substantive model, a Poisson process. Melatos et al. [50] demonstrated that glitch waiting times are consistent with an avalanche process transferring angular momentum from the core superfluid to the crust. Choosing nine pulsars that had glitched five times or more, they found that seven of these were consistent with a constant-rate Poisson process such that each glitch event was statistically independent. In the remaining two, PSR J0537-6910 and PSR B0833-45 (Vela), they found that a quasiperiodic component (i.e., the glitches occur quasiperiodically) coexists with the Poisson process and accounts for about 20% of the events; this is suggestive that these periodic glitches originate from a different mechanism.

Of all events in the glitch catalog, PSR J0537-6910 accounts for 23 and PSR B0833-45 (Vela) for 17 of the total 472 events. Assuming that 20% of these are due to the quasiperiodic component, this is ∼1.7% of the total number of observed glitches. It is possible that other pulsars also exhibit a quasiperiodic component, so the total fraction of glitches from a quasiperiodic component may be larger than 1.7%, but it seems likely that a Poisson-like process should provide a good description of the probability of glitches occurring in general.

Assuming the Poisson process is responsible for all the glitches in the catalog, we can calculate the probability of one or more glitches occurring given the expected number of glitches. To do this, we take the estimated number of glitches during a typical search $\lambda$ and sum the Poisson probability mass function from 1 to infinity:

$$P(N_g \geq 1; \lambda) = \sum_{N_g=1}^{\infty} \frac{\lambda^{N_g} e^{-\lambda}}{N_g!}. \tag{31}$$

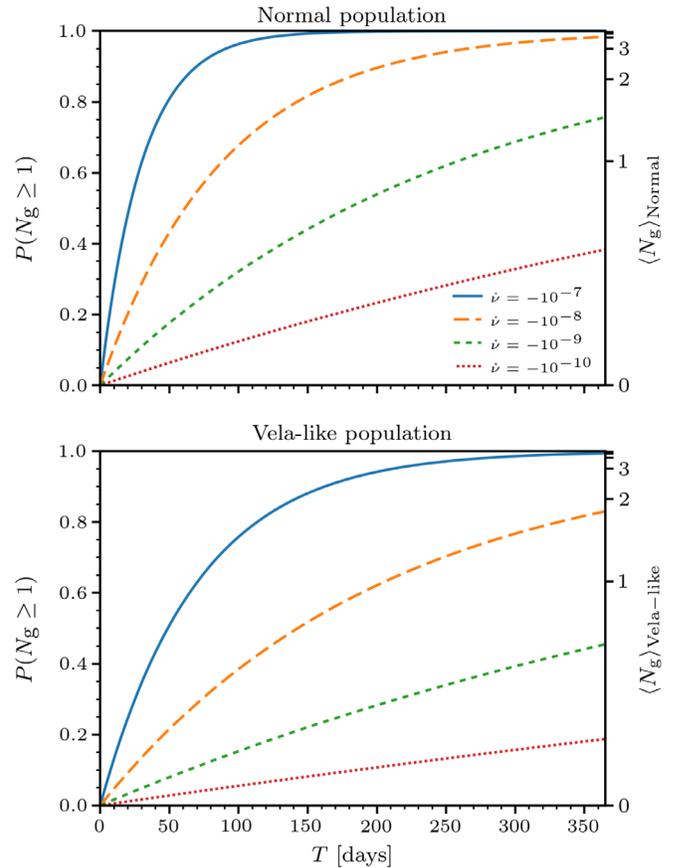

FIG. 8. Probability of one or more glitches occurring during a search of duration $T$, for a normal population (upper plot) and a Vela-like population (lower plot), plotted for four different spin-down rates (in units of Hz/s). The right-hand axis shows the corresponding expected number of glitches.

Note that, in practice, we truncate the summation at a finite level where the mass function is negligible. We will apply this separately to the normal and Vela-like populations with $\lambda = \langle N_g \rangle_{\text{normal}}$ and $\lambda = \langle N_g \rangle_{\text{Vela-like}}$. The total expected number of glitches is the sum of the individual expectations, and from this, the total probability could be calculated.

In Fig. 8, we illustrate how this probability varies with the data span for four fixed spin-down rates. This shows that for the normal population, if the data span is shorter than a month or so, glitches are unlikely even for large (absolute) spin-down rates. However, for more typical data spans $\gtrsim 100$ days, the probability of seeing one glitch rises above one-half with the effect correlating with the magnitude of the spin-down. The same general pattern is found by the Vela-like population, with the effect being marginally weaker due to our prior weighting.

### B. Past and future searches

As discussed, neutron stars with large spin-down rates will have larger and more frequent glitches than those with smaller spin-down rates; see Figs. 4, 5, and 8. It follows that gravitational-wave searches that search over larger spin-down ranges and over longer spans of data will suffer greater mismatches due to glitches than searches over smaller ranges in the spin-down rate and over shorter observation spans. In this section, we aim to quantify the potential mismatch for some recent and ongoing CW searches.

For each CW search that we consider, we estimate the probability of glitches and the mismatch if one or more glitches did occur. We evaluate these quantities for signals with the largest spin-down rate (i.e., largest value of $-\dot{\nu}$) considered in each search. In this respect, it represents a worst-case scenario, but we note that the results would not change much if we had instead averaged over the spin-down rate of the search; averaging uniformly over spin-down would still correspond to the scale set by the largest spin-down rate. These numbers are therefore still qualitatively representative for the "expected" effects over the whole search space. For each search listed in Table III, we summarize the relevant search parameters followed by our estimates for the effect of glitches in this search. Splitting the results up by their source population (such that to get the total expected number of glitches one adds the normal and Vela-like expectations), we list the expected number of glitches, the probability for one or more glitches, and the semicoherent and fully coherent mismatch expectations if one or more glitches did occur.

These mismatches have been computed using an $\mathcal{F}$ statistic-based MCMC search as described in Sec. IV E 2, while the glitch parameters have been sampled as follows: given the expected number of glitches computed at the largest spin-down rate, we draw the actual number of glitches from a Poisson distribution, cf. Eq. (31). If the number of glitches is zero, the mismatch would be zero (since we assume an infinitely fine template bank). In the case of nonzero glitches, we pick the glitch times uniformly throughout the search span. For each glitch, we draw glitch sizes based on a Gaussian distribution with a mean and standard deviation (in log space) obtained from the fitting formulas in Eqs. (1)–(4). For each such glitching signal, we then numerically compute the mismatch using the $\mathcal{F}$-statistic MCMC search described in Sec. IV E 2 over the search dimensions of the original search. Subsequently, we average these mismatches (excluding cases where no glitches occur) to produce the expected mismatch numbers given in Table III.

In brackets, next to the averaged mismatch, we also provide the average upper bound of Eqs. (26) and (28), respectively. This number is computed directly from $\langle N_g \rangle$ as follows. First draw a large sample of random Poisson variables $N_g^i \sim \text{Poisson}(\langle N_g \rangle)$. For each instance $N_g^i$, generate a realization of the fractional time at which the glitches occur (i.e., draw $N_g^i$ uniform random variables from [0, 1]), and record the maximum interglitch fractional duration $\max_a R_{(a)}^i$, which fully determines the mismatch upper bounds. We note that for many cases in this table the average expected mismatch is close to and in some cases even larger than the average upper bound. In the cases in which it is larger than the upper bound, this indicates that our MCMC minimization failed to always find the global minimum. Upon inspection, it appears that this happens most when there is a large number of glitches (as is the case for the S6 E@H Cassiopeia A search and O1 E@H Multi-directed Cassiopeia A). In the cases in which it is close to the upper bound, we conclude that the glitch sizes are so large that the mismatch often saturates at the upper bounds.

TABLE III. Predicted effect of glitches on some recent and planned CW searches; note that the "S6 E@H all-sky HFU*" search refers to stage 4 of the hierarchical follow-up. For each search, we summarize the search parameters, followed by estimates for the number and probability of glitches, and the resulting averaged semicoherent ($\langle \hat{\mu}^{(0)} \rangle$) and fully coherent ($\langle \tilde{\mu}^{(0)} \rangle$) $\mathcal{F}$-statistic mismatches (see the text for details). These quantities are calculated at the largest spin-down rate considered in the search. We also provide the average value of the corresponding mismatch upper bounds (given in brackets); see the text for details.

| | $\min(\dot{f}_s)$ (nHz/s) | $T_{seg}$ (hr) | $N_{seg}$ | $T$ (days) | Ref. | Normal population | | | | Vela-like population | | | |
|---|---|---|---|---|---|---|---|---|---|---|---|---|---|
| | | | | | | $\langle N_g \rangle$ | $P_{N_g \geq 1}$ | $\langle \hat{\mu}^{(0)} \rangle_{N_g \geq 1}$ | $\langle \tilde{\mu}^{(0)} \rangle_{N_g \geq 1}$ | $\langle N_g \rangle$ | $P_{N_g \geq 1}$ | $\langle \hat{\mu}^{(0)} \rangle_{N_g \geq 1}$ | $\langle \tilde{\mu}^{(0)} \rangle_{N_g \geq 1}$ |
| S6 E@H all sky | −2.7 | 60 | 90 | 255 | [51] | 1.1 | 68% | 0.16(0.35) | 0.40(0.54) | 0.5 | 39% | 0.18(0.29) | 0.34(0.47) |
| S6 E@H all-sky HFU* | −2.7 | 280 | 22 | 257 | [26] | 1.1 | 68% | 0.27(0.36) | 0.44(0.53) | 0.5 | 39% | 0.26(0.31) | 0.39(0.47) |
| S6 E@H Cassiopeia A | −106.0 | 140 | 44 | 257 | [22] | 6.3 | 100% | 0.64(0.63) | 0.93(0.84) | 2.7 | 93% | 0.44(0.45) | 0.76(0.66) |
| O1 E@H all sky | −2.6 | 210 | 12 | 105 | [52] | 0.5 | 37% | 0.17(0.33) | 0.25(0.46) | 0.2 | 18% | 0.17(0.30) | 0.25(0.43) |
| O1 E@H Cassiopeia A | −144.0 | 245 | 12 | 122.5 | [52] | 3.5 | 97% | 0.54(0.54) | 0.77(0.71) | 1.5 | 78% | 0.38(0.40) | 0.59(0.56) |
| O1 E@H Vela Junior | −67.9 | 369 | 8 | 123.0 | [52] | 2.5 | 91% | 0.47(0.49) | 0.66(0.64) | 1.1 | 65% | 0.35(0.39) | 0.51(0.52) |
| O1 E@H G357.3 | −29.7 | 489 | 6 | 122.25 | [52] | 1.7 | 81% | 0.41(0.46) | 0.59(0.58) | 0.7 | 51% | 0.34(0.39) | 0.50(0.49) |

This table shows that for these searches considered there is both a substantial probability of glitches occurring, and these glitches would result in a significant fraction of the SNR being lost. The table also reiterates one of the earlier findings shown in Fig. 6: the mismatches will increase in the follow-up. In this case, we have estimates of the mismatch during the initial semicoherent stage and the mismatch if followed up fully coherently. Taking as an example the S6 E@H all-sky search, during the initial stage, only 0.16 of the squared SNR would be lost, and hence the signal (if it where strong enough) might be classified as a candidate. However, if immediately followed up fully coherently, the mismatch may increase to 0.40 such that the signal would appear to be weaker than it should be if it had matched the presumed signal model.

### C. Including the recovery from glitches

In this work, we have used the glitch catalog [27], which provides $\delta\nu$ and $\delta\dot{\nu}$, estimates of the change in rotation frequency and spin-down rate of the pulsar at the glitch. However, in addition to this instantaneous behavior, some pulsars also undergo a short-term exponential relaxation of some fraction $Q$ of the total glitch magnitude over time scales $\tau^d$ that are typically tens to hundreds of days [34]. This may have an important effect on our estimates since, if a large fraction of the glitch is recovered in a time scale that is short compared to the observation time, we will overestimate the mismatch. On the other hand, if not explicitly included in the search template, the relaxation itself could also cause a mismatch, which would tend to vanish when $\tau^d \ll T$ but be maximal when $\tau^d \sim T$.

One can ask what the available pulsar data tell us about the likely values of $Q$. Unfortunately, the issue of extracting the recovery parameter $Q$ from pulsar data is a subtle one. As noted by Ref. [27] in relation to the glitch catalog, "The results presented in this paper do not involve fitting of short-term recoveries because their parameters depend so critically upon the usually poorly known glitch epoch." Furthermore, if the recovery time scale is short compared to the baseline on which pulsar timings are constructed, the recovery will not be apparent and will effectively be absorbed into the measured $\delta\nu$ values. If, instead, the recovery is on a time scale that is long compared to the baseline on which the pulsar timings are made (but not so long that one cannot see the recovery by connecting different timing solutions), then the recovery will be visible in the data, and recovery values $Q$ will be estimated.

Such procedures have been carried out, with recovery values quoted for many glitches, e.g., in the work by Lyne *et al.* [34] and Yu *et al.* [53]. The results of Lyne *et al.* [34] pointed to a correlation between $Q$ and the spin-down rate, with the largest recovery fractions being found in the pulsars with the largest spin-down rates (see their Fig. 6). However, the more recent data of Yu *et al.* [53] point to a more complex picture, with the largest glitches having values of $Q$ from $\sim 10^{-3}$ up to $Q \sim 1$.

Furthermore, a recent detailed study by Shannon *et al.* [54] concluded that eight glitches in the Vela pulsar are best described by permanent and transient changes in the frequency alone, with a dominant decay time scale of 1300 days and $Q \sim 0.3$–$0.8$. If such findings are replicated in many other pulsars, we might need to revisit the modeling of glitches used here, including the transient component but removing the jumps in $\dot{\nu}$; however, in doing so, we would not expect drastic changes to our overall conclusions, since the decay time is long compared to typical observation spans. Given that the glitch catalog modeled glitches with permanent offsets in frequency and derivative only, the analysis performed here is appropriate and consistent with the current literature.

Currently, it is unclear what $Q$ and decay time scale we might expect for the target population of all-sky and directed gravitational-wave searches. It is certainly possible that in the stars of interest to us the recovery fraction may be large ($Q \sim 1$) so that the results of Table III overestimate the effect of glitches on CW searches. However, on the basis of the available evidence, it is not possible to say more than this. The only way of settling the issue definitively would be to allow for relaxation in CW templates and see if this better explains the gravitational-wave data.

## VI. CONCLUSIONS

This work investigates the effects of glitches in the CW signal on searches that matched filter the data against nonglitching templates. We started with an initial study developing empirical fitting formulas for the size and rate of glitches in the pulsar population (as cataloged by Espinoza *et al.* [27]). Subsequently, we developed different ways to estimate and evaluate the loss of SNR in fully coherent and semicoherent searches due to the presence of glitches. Finally, we used our fitting formulas to predict the loss of SNR for typical CW searches and the probability of one or more glitches occurring.

The work is developed with three motivations: (i) to understand if glitches pose a risk to our ability to detect CWs from isolated neutron stars in wide-parameter space searches, (ii) to help guide decisions about planned future searches in order to minimize any risk, and (iii) to make a statement on how the risk manifests and what can be done to mitigate it.

With regard to the first motivation, it is clear from Sec. V that glitches, if they occur in CW signals with properties similar to those observed in pulsars, can both be frequent enough and of a sufficient magnitude to result in large losses of SNR for both semicoherent and fully coherent searches. This is particularly true for searches spanning long stretches of data and searching large spin-down rates, two attributes usually associated with increased detectability.

The main uncertainty in our analysis lies in extrapolating glitch sizes and rates from the known pulsar population to regions of large spin-down rates appropriate to CW searches. In the results presented here, we have had to extrapolate by about 2 orders of magnitude in the spin-down rate $\dot{\nu}$. We have also not attempted to account for postglitch relaxation in our analysis, as there is not a sufficiently clear pattern of how the amount of relaxation correlates with other quantities. We simply note that glitch recovery could serve to significantly reduce the impact of glitches on gravitational-wave searches, if the amount of relaxation is sufficiently large and if it occurs on time scales of relevance to CW searches.

For semicoherent searches, the impact of glitches is reduced by using shorter segment lengths. However, this loss of the SNR is not the primary risk; even if a glitching signal has been identified as a candidate in the initial wide-parameter space search, the greater risk potentially lies in the follow-up procedure of these candidates, which is often considered the test for a real signal [26]. During the follow-up, for standard CW signals, it is expected that the squared SNR grows linearly with increasing segment length [e.g., see Eq. (C7)]. However, for a glitching signal (as seen in Fig. 6) the mismatch will also grow, and so the SNR will not increase in the expected way, and the candidate may potentially be dismissed.

This is of concern to both future and past searches for CWs from neutron stars. If the effect of glitches is ignored, detectable signals could easily be missed due to the presence of glitches. We therefore recommend that the setup design of semicoherent searches and follow-ups take the possibility of glitching signals into account, especially when searching at large spin-down rates. Furthermore, more work is needed to investigate and implement modifications to current follow-up procedures in order to account for the possibility of glitches and ideally be able to fully localize them in CW signals.

## ACKNOWLEDGMENTS

D. I. J. acknowledges support from STFC and also travel support from NewCompStar (a COST-funded Research Networking Programme). All authors are grateful to Christobal Espinoza for help with using the glitch catalog and Danai Antonopoulou, Wynn Ho, Graham Woan, Matt Pitkin, and members of the Continuous Waves group of the LIGO Scientific Collaboration and the Virgo Scientific Collaboration for useful feedback and discussion. This article has been assigned the document number LIGO-P1600201.

## APPENDIX A: BAYESIAN MODEL COMPARISON: TEST OF MIXTURE MODELS

It seems clear by eye that the histogrammed magnitudes of the frequency change in a glitch, $\log_{10}|\delta\nu|$, as shown in Fig. 1, exhibit at least two distinct modes. To model this empirically, we will use a Gaussian mixture model (GMM) [31] with $N$ components. This model assumes that the measured data are taken from a population with $N$ subpopulations, each having a Gaussian distribution with separate mean, variance, and weight ($\mu_i$, $\sigma_i^2$, $\omega_i$), where $i \in [1, N]$; note that $\sum_1^N \omega_i = 1$. Furthermore, we can also allow each of the components to be skewed with a dimensionless skew parameter $\alpha_i$, which can be either positive or negative determining the direction of the skew or 0, for which there is no skew. Following O'Hagan and Leonard [32], then, the probability density function of the $i$th skewed Gaussian component is

$$f(x; \mu_i, \sigma_i, \alpha_i) = 2\mathcal{N}(x; \mu_i, \sigma_i) \int_{-\inf}^{x} \mathcal{N}(\alpha_i x; \mu_i, \sigma_i) dx, \quad (A1)$$

where $\mathcal{N}$ denotes the Gaussian distribution.

Let **y** be the set of measured values of $\log_{10}|\delta\nu|$ and $\vartheta = \{\mu_i, \sigma_i, \alpha_i, \omega_i\}$ be the collection of all model parameters. Then, the probability density for a GMM with $N$ components is

$$P(\mathbf{y}|\text{model}, \vartheta) = \sum_{i=1}^{N} \omega_i f(y_i; \mu_i, \sigma_i, \alpha_i). \quad (A2)$$

To compare different choices of $N$, we will perform a Bayesian model comparison [55] between each of the mixture models and the simplest hypothesis, a mixture model with $N = 1$.

For each model parameter, we must specify a prior. We list these in Eq. (A3), having defined $\langle y \rangle$, $|y|$, and $\text{std}(y)$ as the average, range, and standard deviation of the data,

$$\begin{aligned}
P(\mu_i) &= \text{Unif}(\langle y \rangle - |y|, \langle y \rangle + |y|), \\
P(\sigma_i) &= \text{Half-Cauchy}(0, \text{std}(y)/2), \\
P(\omega_i) &= \text{Unif}(0, 1), \\
P(\alpha_i) &= \mathcal{N}(0, 10 \times \text{std}(y)).
\end{aligned} \quad (A3)$$

For the mean $\mu_i$, we use a uniform prior over a range of values containing all data points. For the standard deviation $\sigma_i$, we will use a half-Cauchy distribution with zero mean as suggested by Gelman *et al.* [56]. A large standard deviation, as compared to the standard deviation of the data themselves, provides a weakly informative prior. Instead, we use a standard deviation of $\frac{\text{std}(y)}{2}$ to favor

TABLE IV. Bayes factor for all models considered in this study compared to the simplest N = 1 GMM. The error is an estimate of the numerical error in the thermodynamic integration.

| Model | $\log_{10}\left(\frac{P(\text{model}|\mathbf{y})}{P(N=1\,\text{GMM}|\mathbf{y})}\right)$ |
|---|---|
| 2 components | $39.12 \pm 0.19$ |
| 2 components (skewed) | $41.60 \pm 0.21$ |
| 3 components | $42.70 \pm 0.23$ |
| 4 components | $44.27 \pm 0.24$ |
| 5 components | $44.18 \pm 0.22$ |
| 6 components | $43.21 \pm 0.22$ |
| 7 components | $42.26 \pm 0.22$ |

GMM components with small standard deviations as compared to the data. That is, our prior disfavors models in which any of the components are wide and flat. The prior for $\omega_i$ is uniform on [0, 1] and for $\alpha_i$ is normally distributed with zero mean and a wide, weakly informative standard deviation. The choice of a zero mean favors nonskewed components. Note that the nonskewed models do not include $\alpha_i$ as a model parameter and the GMM with $N = 1$ does not include $\omega_i$.

We use this choice of prior for the model parameters of each component in the GMM with $N$ components. In this way, models with larger values of $N$ have a larger "prior volume," and hence there is a natural Occam factor favoring the simpler models with fewer components; this prevents overfitting.

We will present results for the Bayes factor between a GMM with $N$ components and the simplest model, a GMM with $N = 1$ components. This is computed by

$$\frac{P(\text{model}|\{y_i\})}{P(N=1|\{y_i\})} = \frac{\int_\vartheta P(\{y_i\}|N\,\text{GMM})P(\vartheta)d\vartheta}{\int_\vartheta P(\{y_i\}|N=1\,\text{GMM})P(\vartheta)d\vartheta}. \quad (A4)$$

We use the EMCEE [57] MCMC algorithm to sample from the posterior and thermodynamic integration to estimate the evidence integrals [58]. In Table IV, we provide the $\log_{10}$ of the Bayes factor for several possible models. The Bayes factor between any two of the models given in Table IV can be calculated from their difference.

This table clearly shows that the data are decisive; a Gaussian mixture model with $N \geq 2$ fits the data a great deal better than the simple $N = 1$ GMM. This is unsurprising, given the distinct multimodal nature of the data. However, the differences between the other models is more subtle. No single model distinguishes itself by a decisive odds ratio compared to its neighboring models. We have checked that these results are robust to small changes in the prior specification.

To help illustrate the differences between these models, in Fig. 9, we plot the probability density for the maximum posterior model parameters found in a few selected models. It is clear from these plots that the $N = 2$ model does not explain the number of glitches found in between the two primary subpopulations around $\log_{10}|\delta\nu/\text{Hz}| = -5.5$; by comparison, the $N > 2$ models and the $N = 2$ model that allows for skewness can explain these points, and this is reflected in the Bayes factor.

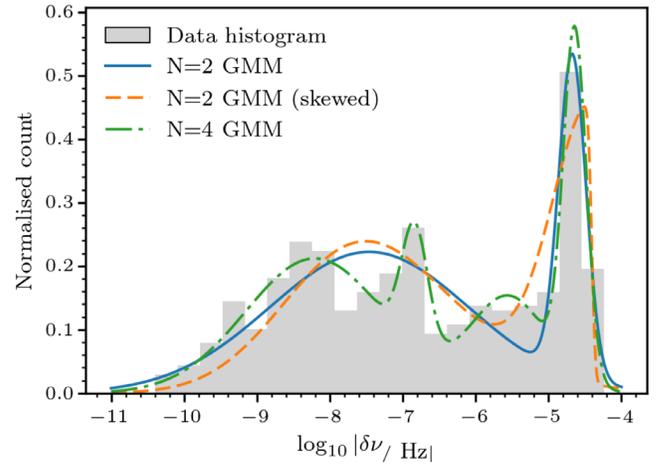

FIG. 9. The distribution of glitch sizes in frequency jump along with the predictions for the components of several GMM; the Bayes factor for these models can be found in Table IV.

From this analysis, it is difficult to decide which model best fits the data. However, what is clear is that simply modeling the data as a GMM with two components with a skew provides a reasonable empirical model. For this reason, in our analysis of the glitch population, we will use this model and not any of the models with a greater number of components.

It is important to realize that this comparison is purely empirical, in that the result was not conditioned on a substantive physical model. It would be interesting to include such modeling; this may provide some insight into the appropriateness of the mixture model and the number of components.

## APPENDIX B: LINEAR REGRESSION IN LOG SPACE

In Sec. III, we perform several linear regressions in log space in order to calculate power-law fits. This assumes that the observed values $\log(y_i)$ depend on the predictor values $\log(x_i)$ as

$$\log(y_i) = m\log(x_i) + c + \epsilon_i, \quad (B1)$$

where the $\epsilon_i$ are independent and identically central normally distributed variables with a standard deviation $\sigma$. In this way, $m$ and $c$ are the linear fit free variables, while $\sigma$ is a measure of the variability in the observation about this linear fit.

We use a Bayesian linear regression in which we estimate the posterior distributions of all three parameters using a Markov-chain Monte Carlo algorithm; for the prior distributions, we use noninformative priors and test that these do not induce any bias. In all cases, we find the resulting posteriors to be Gaussian and so can take their mean values to get best-fit parameters. The advantage of this method compared to a simple least-squares linear regression is that we also estimate $\langle\sigma\rangle$, the variation about the linear fit. The linear fit can therefore be written as

$$\log(y(x)) = \langle m\rangle \log(x) + \langle c\rangle \pm \langle\sigma\rangle. \quad (B2)$$

We can then rearrange this equation to give the corresponding power-law fit in linear space,

$$y(x) = 10^{\langle c\rangle} x^{\langle m\rangle} 10^{\pm\langle\sigma\rangle}, \quad (B3)$$

where the last term gives the variability about the mean. Hence, neglecting this term gives the mean.

This is an inherently problematic approach since many functions besides a power law can appear linear in a log-log plot and the assumption of Gaussian error may not be a good description. Nevertheless, we will still apply this approach since we need only order-of-magnitude estimates and can always check our predictions; we must be clear that the power-law fit gives a good empirical fit but is not intended to signify any substantive underlying model.

## APPENDIX C: DERIVATION OF GLITCH MISMATCHES

### 1. Coherent matched-filtering approximation

To study glitch mismatches, it is useful to review the (simplified) matched-filtering amplitude introduced by Prix and Itoh [59], from which the metric can easily be derived. The coherent matched-filter scalar product for narrow-band, long-lasting signals can be written [45] as

$$(x|y) = \frac{2T}{S_h}\langle xy\rangle, \quad (C1)$$

in terms of the time average

$$\langle Q\rangle \equiv \frac{1}{T}\int_{t_0}^{t_0+T} Q(t)dt, \quad (C2)$$

where $t_0$ denotes the start time and $T$ denotes the duration of the data span analyzed. The log-likelihood ratio $\ln\Lambda$ of the signal model vs Gaussian noise is found as (e.g., see Ref. [5])

$$\ln\Lambda(x;\theta) = (x|h(\theta)) - \frac{1}{2}(h(\theta)|h(\theta)), \quad (C3)$$

where $x(t)$ is the data time series and $h(t;\theta)$ is a template waveform with phase parameters $\theta$. To derive the phase metric, it is useful to consider a simplified constant-amplitude signal model of the form

$$h(t;\theta) = A\cos\Phi(t - t_{\text{ref}};\theta). \quad (C4)$$

We note that the dominant term in the phase function is $\Phi(t) \propto 2\pi ft$, where typical signal frequencies in ground-based detectors are $f \gtrsim 10$ Hz and typical observation times $T$ are of order of a day or longer, which results in

$$(h|h) \approx \frac{T}{S_h} AT^2. \quad (C5)$$

Assuming the data to be dominated by a signal (neglecting noise), i.e., $x(t) \approx s(t) = A_s\cos\Phi(t - t_{\text{ref}};\theta_s)$, and maximizing the log-likelihood ratio $\ln\Lambda$ over the template amplitude $A$ [but *not* yet over the initial phase $\phi \equiv \Phi(0;\theta)$, which, contrary to the standard treatment, is left as an explicit template parameter for now] yields the (partial) maximum-likelihood expression

$$\ln\Lambda_{\text{ML}_A}(\theta) = \frac{2A_s^2 T}{S_h}\langle\cos\Phi_s\cos\Phi_t\rangle^2$$
$$\approx \frac{1}{2}\rho_s^2[\Re X(\theta_s;\theta)]^2, \quad (C6)$$

with the perfect-match signal squared SNR defined as

$$\rho_s^2 \equiv (s|s) \approx \frac{A_s^2 T}{S_h} \quad (C7)$$

and the (coherent) "matched-filtering amplitude" $X$ defined as

$$X(\theta_s;\theta) \equiv \langle e^{i\Delta\Phi(\theta_s;\theta)}\rangle, \quad (C8)$$

with $|X|^2 \leq 1$ and where $\Delta\Phi \equiv \Phi(t;\theta_s) - \Phi(t;\theta)$. Note that, contrary to the discussion by Prix and Itoh [59], here $X$ still depends on the initial-phase parameter $\phi$, and one can verify that maximizing $X$ over $\phi$ yields again the expression found by Prix and Itoh [59], namely, $\ln\Lambda_{\text{ML}_{A,\phi}} = \frac{1}{2}\rho_s^2|X|^2$.

From Eq. (C6), it is natural to define the "mismatched ($\phi$-coherent) SNR" $\breve{\rho}(\theta_s;\theta)$ as

$$\breve{\rho}^2(\theta_s;\theta) \equiv \rho_s^2[\Re X(\theta_s;\theta)]^2, \quad (C9)$$

and plugging this into the mismatch definition of Eq. (7) yields the $\phi$-coherent mismatch

$$\breve{\mu}^{(0)}(\theta_s;\theta) = 1 - [\Re X(\theta_s;\theta)]^2. \quad (C10)$$

Specializing this to the glitching signal model of Eq. (11), we obtain

$$X = \sum_{a=0}^{N_g} R_{(a)} X_{(a)}, \quad \text{with}$$

$$R_{(a)} \equiv \frac{T_{(a)}}{T}, \quad \text{and}$$

$$X_{(a)} \equiv \frac{1}{T_{(a)}} \int_{t_{(a)}}^{t_{(a+1)}} \exp\{i[\Phi(t;\theta_{(a)}) - \Phi(t;\theta)]\}. \quad \text{(C11)}$$

Taylor expanding to second order around the perfect-match case $\Delta\theta_{(a)} = 0$ in each continuous stretch yields

$$\Re X_{(a)}(\theta_s; \Delta\theta_{(a)}) \approx 1 - \frac{1}{2} \breve{g}_{\alpha\beta}^{(a)} \Delta\theta_{(a)}^{\alpha} \Delta\theta_{(a)}^{\beta}, \quad \text{(C12)}$$

with implicit summation over repeated phase-parameter indices $\alpha$, $\beta$, and with the $\phi$-coherent metric $\breve{g}_{\alpha\beta}^{(a)}$ for stretch $a$ defined as

$$\breve{g}_{\alpha\beta}^{(a)} \equiv \frac{1}{T_{(a)}} \int_{t_{(a)}}^{t_{(a+1)}} \partial_\alpha \Phi \partial_\beta \Phi \, dt. \quad \text{(C13)}$$

This "$\phi$-coherent" form of the phase metric differs from the standard "coherent" expression (e.g., see the work by Prix [38] and Brady *et al.* [41]) as it still covers the initial-phase parameter $\phi$. Minimizing the $\phi$-coherent mismatch over $\phi$, one can easily recover the usual form of the standard coherent metric, namely,

$$\tilde{g}_{ij} = \langle \partial_i \Phi \partial_j \Phi \rangle - \langle \partial_i \Phi \rangle \langle \partial_j \Phi \rangle, \quad \text{(C14)}$$

where indices $i, j = 1, \ldots$ run over the phase parameters excluding $\phi$, while indices $\alpha$, $\beta$ label all phase parameters including $\phi$.

Substituting Eq. (C12) into Eqs. (C10) and (C11) and keeping only terms up to second order in $\Delta\theta_{(a)}$, we obtain the $\phi$-coherent metric mismatch approximation as

$$\breve{\mu} = \sum_{a=0}^{N_g} R_{(a)} \breve{g}_{\alpha\beta}^{(a)} \Delta\theta_{(a)}^{\alpha} \Delta\theta_{(a)}^{\beta}. \quad \text{(C15)}$$

### 2. Minimal coherent metric mismatch

We can use the general $\phi$-coherent multiglitch mismatch expression of Eq. (C15) to express the minimal mismatch over an infinitely fine template bank of Eq. (8). For this purpose, we parametrize the offsets $\Delta\theta_{(a)}$ of Eq. (15) between the signal and template in terms of the offset in the first continuous stretch $a = 0$ before the first glitch; i.e., we write

$$\Delta\theta_{(0)} \equiv \Delta\theta, \quad \text{(C16)}$$

$$\Delta\theta_{(a)} = \Delta\theta + \delta\theta_{(a)} \quad \text{for } a \geq 1, \quad \text{(C17)}$$

and so we can minimize the mismatch over the template bank by varying $\Delta\theta$. From Eq. (C15), we obtain

$$\breve{\mu} = \breve{g}_{\alpha\beta} \Delta\theta^{\alpha} \Delta\theta^{\beta} + 2 \sum_{a=1}^{N_g} R_{(a)} \breve{g}_{\alpha\beta}^{(a)} \delta\theta_{(a)}^{\alpha} \Delta\theta^{\beta}$$
$$+ \sum_{a=1}^{N_g} R_{(a)} \breve{g}_{\alpha\beta}^{(a)} \delta\theta_{(a)}^{\alpha} \delta\theta_{(a)}^{\beta}, \quad \text{(C18)}$$

where $\breve{g} \equiv \sum_{a=0}^{N_g} \breve{g}^{(a)}$ is the $\phi$-coherent metric for non-glitching signals over the whole duration $T$. Minimizing Eq. (C18) over $\Delta\theta$ by solving $\partial\breve{\mu}/\partial\Delta\theta^{\alpha} = 0$ yields the minimizing template offset as

$$\Delta\theta_{\min}^{\alpha} = -\breve{g}^{\alpha\gamma} \sum_{a=1}^{N_g} R_{(a)} \breve{g}_{\gamma\delta}^{(a)} \delta\theta_{(a)}^{\delta}, \quad \text{(C19)}$$

where we denoted the inverse $\phi$-coherent metric for non-glitching signals as $\breve{g}^{\alpha\beta} \equiv \{\breve{g}^{-1}\}^{\alpha\beta}$. Inserting this into Eq. (C18) yields the minimal *coherent* glitch mismatch as

$$\tilde{\mu}_{\min} = \sum_{a,b=1}^{N_g} \delta\theta_{(a)}^{\alpha} \tilde{G}_{\alpha\beta}^{(a)(b)} \delta\theta_{(b)}^{\beta}, \quad \text{(C20)}$$

in terms of

$$\tilde{G}_{\alpha\beta}^{(a)(b)} \equiv \delta_{ab} R_{(a)} \breve{g}_{\alpha\beta}^{(a)} - R_{(a)} R_{(b)} \breve{g}_{\alpha\gamma}^{(a)} \breve{g}^{\gamma\delta} \breve{g}_{\delta\beta}^{(b)}. \quad \text{(C21)}$$

### 3. Coherent metric mismatch for a single glitch

In the special case of a single glitch $N_g = 1$, we write $\delta\theta_{(1)} \equiv \delta\theta$ and $R \equiv R_{(0)} = T_{(0)}/T$ and therefore $R_{(1)} = 1 - R$, and $T_{(1)} = (1 - R)T$, and so Eqs. (C20) and (C21) yield

$$\tilde{\mu}_{\min} = \delta\theta^{\alpha} \tilde{G}_{\alpha\beta} \delta\theta^{\beta}, \quad \text{with} \quad \text{(C22)}$$

$$\tilde{G}_{\alpha\beta} = (1 - R) \breve{g}_{\alpha\beta}^{(1)} - (1 - R)^2 \breve{g}_{\alpha\gamma}^{(1)} \breve{g}^{\gamma\delta} \breve{g}_{\delta\beta}^{(1)}, \quad \text{(C23)}$$

and assuming the glitch occurs with uniform probability in $R \in [0, 1]$, the expected mismatch for a single glitch can be found as

$$\langle \tilde{\mu}_{\min} \rangle_R = \delta\theta^{\alpha} \left[ \frac{1}{2} \breve{g}_{\alpha\beta}^{(1)} - \frac{1}{3} \breve{g}_{\alpha\gamma}^{(1)} \breve{g}^{\gamma\delta} \breve{g}_{\delta\beta}^{(1)} \right] \delta\theta^{\beta}. \quad \text{(C24)}$$

To obtain an explicit estimate for the minimal expected glitch mismatch of Eq. (C24), we consider the case of a "directed" search for a target with known sky position and binary-orbital parameters, where the search space only includes frequency $f$ and spin-downs $\{\dot{f}, \ddot{f}, \ldots\}$. Note, however, that generally there will be metric correlations also between sky-position or binary-orbital phase parameters and the set of "glitch parameters" $\{\phi, f, \dot{f}\}$, which can result in further reductions in the minimal mismatch of Eq. (C20) compared to the example given here.

Assuming a directed search, the phase model is simply

$$\Phi(\Delta t; \theta) = \phi + 2\pi \left[ f\Delta t + \frac{1}{2}\dot{f}\Delta t^2 + \frac{1}{3!}\ddot{f}\Delta t^3 + \ldots \right], \quad (C25)$$

where $\Delta t = t - t_{\text{ref}}$ is the offset from the reference time $t_{\text{ref}}$ at which the phase parameters $\{\phi, f, \dot{f}, \ddot{f}, \ldots\}$ are defined. It is easy to see that changing the reference time to $t'_{\text{ref}} = t_{\text{ref}} + \tau$ results in new coordinates

$$\theta^{\alpha'} \equiv \theta^{\alpha}(t_{\text{ref}} + \tau) = \mathcal{T}^{\alpha}{}_{\beta}(\tau)\theta^{\beta}(t_{\text{ref}}), \quad (C26)$$

with the reference-time shift operator $\mathcal{T}(\tau)$ given by

$$\mathcal{T}(\tau) = \begin{pmatrix} 1 & 2\pi\tau & \pi\tau^2 & \frac{\pi}{3}\tau^3 & \ldots \\ 0 & 1 & \tau & \frac{1}{2}\tau^2 & \ldots \\ 0 & 0 & 1 & \tau & \ldots \\ 0 & 0 & 0 & 1 & \ldots \\ \vdots & \vdots & \vdots & \vdots & \ddots \end{pmatrix}, \quad (C27)$$

which moves the reference time forward in time by $\tau$ and which has the property $\mathcal{T}^{-1}(\tau) = \mathcal{T}(-\tau)$. Note that the functional form of Eq. (C25) remains unchanged under a change of reference time, which simply takes the form of a Taylor expansion around the new reference time, with corresponding Taylor coefficients $\theta'$. We therefore obtain the reference-time shift operation on the metric, as the metric mismatch $\mu$ is invariant under changes of reference time, and so

$$\mu = \Delta\theta^{\alpha}(t_{\text{ref}})\breve{g}_{\alpha\beta}(t_{\text{ref}})\Delta\theta^{\beta}(t_{\text{ref}})$$
$$= \Delta\theta^{\alpha}(t_{\text{ref}} + \tau)\breve{g}_{\alpha\beta}(t_{\text{ref}} + \tau)\Delta\theta^{\beta}(t_{\text{ref}} + \tau)$$
$$= \Delta\theta^{\alpha}(t_{\text{ref}})[\mathcal{T}^{\gamma}{}_{\alpha}(\tau)\breve{g}_{\gamma\delta}(t_{\text{ref}} + \tau)\mathcal{T}^{\delta}{}_{\beta}(\tau)]\Delta\theta^{\beta}(t_{\text{ref}}), \quad (C28)$$

from which we obtain

$$\breve{g}_{\alpha\beta}(t_{\text{ref}} + \tau) = \mathcal{T}^{\gamma}{}_{\alpha}(-\tau)\breve{g}_{\gamma\delta}(t_{\text{ref}})\mathcal{T}^{\delta}{}_{\beta}(-\tau). \quad (C29)$$

This relation is very useful as the metric of Eq. (C13) takes the simplest form when evaluated for a reference time in the middle of the time segment $T$ for which the metric is computed; namely, for $t_{\text{ref}} = t_0 + T/2$, we find

$$\breve{g}_{\text{mid}} = \begin{pmatrix} 1 & 0 & \frac{\pi T^2}{12} & 0 & \ldots \\ 0 & \frac{\pi^2 T^2}{3} & 0 & \frac{\pi^2 T^4}{120} & \ldots \\ \frac{\pi T^2}{12} & 0 & \frac{\pi^2 T^4}{80} & 0 & \ldots \\ 0 & \frac{\pi^2 T^4}{120} & 0 & \frac{\pi^2 T^6}{4032} & \ldots \\ \vdots & \vdots & \vdots & \vdots & \ddots \end{pmatrix}. \quad (C30)$$

For evaluating the mismatch of Eq. (C24), however, it will be most convenient to choose the reference time at the time of the glitch, i.e., $t_{\text{ref}} = t_{(1)}$, such that $\delta\theta$ refers directly to the instantaneous changes in parameters at time $t_{(1)}$. We can compute the corresponding metric simply by appropriately shifting the reference time of Eq. (C30), namely,

$$\breve{g}^{(0)}(t_{(1)}) = \mathcal{T}^{\intercal}(\tau_0) \cdot \breve{g}^{(0)}_{\text{mid}} \cdot \mathcal{T}(\tau_0),$$
$$\breve{g}^{(1)}(t_{(1)}) = \mathcal{T}^{\intercal}(\tau_1) \cdot \breve{g}^{(1)}_{\text{mid}} \cdot \mathcal{T}(\tau_1) \quad (C31)$$

with

$$\tau_0 \equiv -\frac{RT}{2}, \qquad \tau_1 \equiv \frac{(1-R)T}{2}. \quad (C32)$$

Combining these expressions and substituting into Eq. (C23) yields the following $R$-dependent expressions for the mismatch:

$$\tilde{\mu}^{\text{4D}}_{\text{min}}(R) = \delta\phi^2 R(1-R)(1764R^8 - 7056R^7 + 11704R^6 - 10416R^5 + 5376R^4 - 1624R^3 + 276R^2 - 24R + 1)$$
$$+ \delta f^2 T^2 \frac{\pi^2}{3} R^3 (1-R)^3 (588R^6 - 1764R^5 + 2100R^4 - 1260R^3 + 399R^2 - 63R + 4)$$
$$+ \delta\dot{f}^2 T^4 \frac{\pi^2}{45} R^5 (1-R)^5 (180R^4 - 360R^3 + 260R^2 - 80R + 9)$$
$$+ \delta\phi\delta f T 2\pi R^2 (1-R)^2 (2R-1)(7R^2 - 7R + 1)(42R^4 - 84R^3 + 56R^2 - 14R + 1)$$
$$+ \delta\phi\delta\dot{f} T^2 \frac{2\pi}{3} R^3 (1-R)^3 (252R^6 - 756R^5 + 890R^4 - 520R^3 + 156R^2 - 22R + 1)$$
$$+ \delta f\delta\dot{f} T^3 \frac{\pi^2}{3} R^4 (1-R)^4 (2R-1)(6R^2 - 6R + 1)(14R^2 - 14R + 3) \quad (C33)$$

$$\tilde{\mu}_{\min}^{3D}(R) = \delta\phi^2 R(1-R)(175R^6 - 525R^5 + 615R^4 - 355R^3 + 105R^2 - 15R + 1)$$
$$+ \delta f^2 T^2 \frac{4\pi^2}{3} R^3(1-R)^3(21R^4 - 42R^3 + 30R^2 - 9R + 1)$$
$$+ \delta \dot{f}^2 T^4 \frac{\pi^2}{45} R^5(1-R)^5(35R^2 - 35R + 9) + \delta\phi\delta f T 2\pi R^2(1-R)^2(2R-1)$$
$$\times (5R^2 - 5R + 1)(7R^2 - 7R + 1) + \delta\phi\delta\dot{f} T^2 \frac{2\pi}{3} R^3 (1-R)^3$$
$$\times (35R^4 - 70R^3 + 48R^2 - 13R + 1) + \delta f \delta \dot{f} T^3 \frac{\pi^2}{3} R^4 (1-R)^4$$
$$\times (2R-1)(14R^2 - 14R + 3) \tag{C34}$$

$$\tilde{\mu}_{\min}^{2D}(R) = \delta\phi^2 R(1-R)(20R^4 - 40R^3 + 28R^2 - 8R + 1) + \delta f^2 T^2 \frac{\pi^2}{3} R^3(1-R)^3(15R^2 - 15R + 4)$$
$$+ \delta \dot{f}^2 T^4 \frac{\pi^2}{5} R^5(1-R)^5 + \delta\phi\delta f T 2\pi R^2(1-R)^2(2R-1)(5R^2 - 5R + 1)$$
$$+ \delta\phi\delta\dot{f} T^2 \frac{2\pi}{3} R^3(1-R)^3(6R^2 - 6R + 1) + \delta f \delta\dot{f} T^3 \pi^2 R^4 (1-R)^4 (2R-1). \tag{C35}$$

Averaging this over $R$ or directly evaluating Eq. (C24) yields the explicit $R$-averaged metric mismatch expressions given in Eqs. (16)–(18).

### 4. Semicoherent glitch mismatch

Here, the coherent SNR would have been maximized independently over the initial phase $\phi$ in each segment $\ell$ separately before adding, which means that, contrary to the $\phi$-coherent case of the previous section using the full phase parameters $\theta = \{\phi, f, \dot{f}, \ldots\}$, here, we are dealing with the usual coherent statistic in each segment, using the subset of "phase-evolution" parameters

$$\lambda = \{f, \dot{f}, \ddot{f}, \ldots\}, \tag{C36}$$

and we use parameter indices $i, j = 1, 2, \ldots$ to label the components of $\lambda$. Using Eq. (C9), we can obtain the per-segment phase-maximized coherent SNR of segment $\ell$ as

$$\tilde{\rho}_\ell^2(\lambda_s; \lambda) = \tilde{\rho}_s^2 |X_\ell(\lambda_s; \lambda)|^2, \tag{C37}$$

and the expectation of the resulting incoherent statistic is simply given by summing this over all segments, i.e.,

$$\hat{\rho}^2(\lambda_s; \lambda) = \sum_{\ell=1}^{N_{\text{seg}}} \tilde{\rho}_\ell^2(\lambda_s; \lambda). \tag{C38}$$

The corresponding mismatch of Eq. (7) is now obtained as[5]

---
[5]Note that $\hat{\rho}$ is *not* the incoherent SNR but is proportional to it, which is sufficient for the mismatch definition of Eq. (7) to apply as mentioned earlier.

$$\hat{\mu}^{(0)} = \frac{\hat{\rho}_s^2 - \hat{\rho}^2}{\hat{\rho}_s^2}$$
$$= 1 - \frac{1}{N_{\text{seg}}} \sum_{\ell=1}^{N_{\text{seg}}} |X_\ell(\lambda_s; \lambda)|^2$$
$$= \frac{1}{N_{\text{seg}}} \sum_{\ell=1}^{N_{\text{seg}}} \tilde{\mu}^{(0)}, \tag{C39}$$

where we assumed a constant per-segment perfect-match SNR $\tilde{\rho}_s$. By Taylor expanding in small offsets $\lambda = \lambda_s + \Delta\lambda$ around the signal location $\lambda_s$, using Eq. (12) and dropping terms of higher order, we obtain the incoherent metric mismatch approximation as

$$\hat{\mu} = \frac{1}{N_{\text{seg}}} \sum_{\ell=1}^{N_{\text{seg}}} \tilde{g}_{ij}^\ell \Delta\lambda_\ell^i \Delta\lambda_\ell^j, \tag{C40}$$

where $\tilde{g}_{ij}^\ell$ is the usual coherent metric (i.e., minimized mismatch over $\phi$) for segment $\ell$, as given in Eq. (C14).

We consider again a signal that undergoes $N_g$ glitches at times $t_{(a)}$ with $a = 1, \ldots N_g$, and for simplicity, we assume these glitches to fall at or near segment boundaries, i.e., $t_{(a)} \approx \ell_{(a)} T_{\text{seg}}$. Note that if a glitch happens inside a segment $\ell$, the mismatch $\tilde{\mu}_\ell^{(0)}$ is bounded within [0, 1] and at most contributes $1/N_{\text{seg}}$ to the total mismatch. One can always derive an upper bound on the mismatch by effectively removing the affected segment, assuming the glitch to happen in between segments, and adding $1/N_{\text{seg}}$ to the final mismatch. In this section, we will mostly consider the large-$N_{\text{seg}}$ limit, and so this correction will be neglected.

We denote as $N_{(a)}$ the number of segments between glitch $\ell_{(a)}$ and $\ell_{(a+1)} = \ell_{(a)} + N_{(a)}$, with $l_{(0)} = 1$ and $\ell_{(N_g+1)} = N_{\text{seg}}$, and $R_{(a)} = N_{(a)}/N_{\text{seg}}$, and $\Delta\lambda_{(a)}$ are the parameter offsets in the time stretch following glitch $a$.

By regrouping the expression of Eq. (C40) over the interglitch stretches $a$, we obtain

$$\hat{\mu} = \sum_{a=0}^{N_g} R_{(a)} \hat{g}_{ij}^{(a)} \Delta\lambda_{(a)}^i \Delta\lambda_{(a)}^j, \quad (C41)$$

where we defined

$$\hat{g}_{ij}^{(a)} \equiv \frac{1}{N_{(a)}} \sum_{\ell=\ell_{(a)}}^{\ell_{(a)}+N_{(a)}} \tilde{g}_{ij}^{\ell}. \quad (C42)$$

Note that Eq. (C41) is formally identical to the coherent case of Eq. (C15), and when minimizing this over the template bank, we therefore obtain the analogous result to Eq. (C20), namely,

$$\hat{\mu}_{\min} = \sum_{a,b=1}^{N_g} \delta\lambda_{(a)}^i \hat{G}_{ij}^{(a)(b)} \delta\lambda_{(b)}^j, \quad (C43)$$

where we defined

$$\hat{G}_{ij}^{(a)(b)} \equiv \delta_{ab} R_{(a)} \hat{g}_{ij}^{(a)} - R_{(a)} R_{(b)} \hat{g}_{im}^{(a)} \hat{g}^{mn} \hat{g}_{nj}^{(b)}, \quad (C44)$$

with

$$\hat{g}_{ij} = \sum_{a=0}^{N_g} R_{(a)} \hat{g}_{ij}^{(a)} = \frac{1}{N_{\text{seg}}} \sum_{\ell=1}^{N_{\text{seg}}} \tilde{g}_{ij}^{\ell}, \quad \text{and} \quad (C45)$$

$$\hat{g}^{ij} = \{\hat{g}^{-1}\}^{ij}, \quad (C46)$$

in perfect analogy to the coherent case of Appendix C 2.

Considering the single-glitch case of Appendix C 3, we can again derive explicit mismatch expressions. Using the large-$N_{\text{seg}}$ limit, we find the $R$-dependent mismatch expressions as

$$\hat{\mu}_{\min}^{4D}(R) = \delta f^2 T_{\text{seg}}^2 \frac{\pi^2}{3} R(1-R)(175R^6 - 525R^5 + 615R^4 - 355R^3 + 105R^2 - 15R + 1)$$
$$+ \delta\dot{f}^2 N_{\text{seg}}^2 T_{\text{seg}}^4 \frac{\pi^2}{9} R^3(1-R)^3(21R^4 - 42R^3 + 30R^2 - 9R + 1)$$
$$- \delta f \delta\dot{f} N_{\text{seg}} T_{\text{seg}}^3 \frac{\pi^2}{3} R^2(1-R)^2(2R-1)(5R^2 - 5R + 1)(7R^2 - 7R + 1) \quad (C47)$$

$$\hat{\mu}_{\min}^{3D}(R) = \delta f^2 T_{\text{seg}}^2 \frac{\pi^2}{3} R(1-R)(20R^4 - 40R^3 + 28R^2 - 8R + 1) + \delta\dot{f}^2 N_{\text{seg}}^2 T_{\text{seg}}^4 \frac{\pi^2}{36} R^3(1-R)^3(15R^2 - 15R + 4)$$
$$- \delta f \delta\dot{f} N_{\text{seg}} T_{\text{seg}}^3 \frac{\pi^2}{3} R^2(1-R)^2(2R-1)(5R^2 - 5R + 1) \quad (C48)$$

$$\hat{\mu}_{\min}^{2D}(R) = +\delta f^2 T_{\text{seg}}^2 \frac{\pi^2}{3} R(1-R)(3R^2 - 3R + 1) + \delta\dot{f}^2 N_{\text{seg}}^2 T_{\text{seg}}^4 \frac{\pi^2}{9} R^3(1-R)^3$$
$$- \delta f \delta\dot{f} N_{\text{seg}} T_{\text{seg}}^3 \frac{\pi^2}{3} R^2(1-R)^2(2R-1), \quad (C49)$$

while the $R$-averaged mismatch expressions are given in Eqs. (21)–(23).